\documentclass[submission,copyright,creativecommons,colorlinks]{eptcs}
\providecommand{\thisvolume}[1]{this volume of EPTCS, Open Publishing Association}
\providecommand{\event}{GCM 2024} 

\usepackage{xspace}
\usepackage[T1]{fontenc}        
\usepackage[utf8]{inputenc}
\usepackage[inline]{enumitem}
	\setlist{itemsep=0pt,topsep=.5ex}
\usepackage{fontawesome}
\usepackage{booktabs}
\usepackage{tikz}                     
  \usetikzlibrary{automata,matrix,calc,chains,backgrounds,positioning
                 ,fit,decorations.pathmorphing,shapes
                 ,arrows,arrows.meta}
\usepackage{tikz-bpmn}
\usepackage{amstext}
\usepackage{amsthm} 
\usepackage{amsmath}
\usepackage{amssymb}
\usepackage{stmaryrd}
\usepackage{wasysym}
\usepackage{mathtools}
\usepackage{pgfplots}
\usepackage[ruled,vlined,linesnumbered]{algorithm2e}
\usepackage{verbatim}

\pgfplotsset{compat=1.17}
\newcommand{\GrappaRE}{\textit{Grappa~RE}\xspace}
\newcommand{\Grappa}{\textsc{Grappa}\xspace}
\newcommand{\Spikes}{\emph{Spikes}\xspace}
\newcommand{\Pali}{\emph{Palindromes}\xspace}
\newcommand{\Wheels}{\emph{Wheels}\xspace}
\newcommand{\Java}{\textsc{Java}\xspace}
\newcommand{\Json}{\textsc{Json}\xspace}
\newcommand{\Brics}{\texttt{dk.brics.auto\-maton}\xspace}
\newcommand{\Graphstream}{\textsc{GraphStream}\xspace}

\newcommand{\ms}[1]{$#1\,\textrm{ms}$}
\newcommand{\tableref}[1]{\hyperref[#1]{Table~\ref*{#1}}}
\newcommand{\figref}[1]{\hyperref[#1]{Figure~\ref*{#1}}}
\newcommand{\figsref}[1]{\hyperref[#1]{Figures~\ref*{#1}}}
\newcommand{\figandfigref}[2]%
	{\hyperref[#1]{Figures~\ref*{#1}} \hyperref[#1]{and~\ref*{#2}}}
\newcommand{\assref}[2]%
	{\hyperref[#1]{Assumption~\ref*{#1}}.\hyperref[#1]{\ref*{#2}}}
\newcommand{\assrefs}[3]%
	{\hyperref[#1]{Assumptions~\ref*{#1}}.\hyperref[#1]{\ref*{#2}}--\hyperref[#1]{\ref*{#1}}.\hyperref[#1]{\ref*{#3}}}
\newcommand{\tabref}[1]{\hyperref[#1]{Table~\ref*{#1}}}
\newcommand{\sectref}[1]{\hyperref[#1]{Section~\ref*{#1}}}
\newcommand{\sectsref}[1]{\hyperref[#1]{Sections~\ref*{#1}}}
\newcommand{\stepref}[1]{\hyperref[#1]{Step~\ref*{#1}}}
\newcommand{\defref}[1]{\hyperref[#1]{Definition~\ref*{#1}}}
\newcommand{\thmref}[1]{\hyperref[#1]{Theorem~\ref*{#1}}}
\newcommand{\lemmaref}[1]{\hyperref[#1]{Lemma~\ref*{#1}}}
\newcommand{\factref}[1]{\hyperref[#1]{Fact~\ref*{#1}}}
\newcommand{\factsubref}[2]%
	{\hyperref[#1]{Fact~\ref*{#1}}.\hyperref[#1]{\ref*{#2}}}
\newcommand{\exref}[1]{\hyperref[#1]{Example~\ref*{#1}}}
\newcommand{\exsref}[1]{\hyperref[#1]{Examples~\ref*{#1}}}
\newcommand{\algref}[1]{\hyperref[#1]{Algorithm~\ref*{#1}}}
\newcommand{\alineref}[1]{\hyperref[#1]{Line~\ref*{#1}}}
\newcommand{\alinesref}[1]{\hyperref[#1]{Lines~\ref*{#1}}}
\newcommand{\remref}[1]{\hyperref[#1]{Remark~\ref*{#1}}}
\newcommand{\condref}[1]{\hyperref[#1]{Condition~\ref*{#1}}}
\newcommand{\Eqref}[1]{\hyperref[#1]{Equation~\ref*{#1}}}
\newcommand{\constrref}[1]{\hyperref[#1]{Construction~\ref*{#1}}}
\newcommand{\obsref}[1]{\hyperref[#1]{Observation~\ref*{#1}}}
\newcommand{\corref}[1]{\hyperref[#1]{Corollary~\ref*{#1}}}

\newtheoremstyle{plainbreak}%
  {}{}%
  {\itshape}{}%
  {\bfseries}{}
  {\newline}{}
\newtheoremstyle{definitionbreak}%
  {}{}%
  {}{}%
  {\bfseries}{}
  {\newline}{}
\newtheoremstyle{remarkbreak}%
  {}{}%
  {}{}%
  {\itshape}{}
  {\newline}{}
\theoremstyle{definition}

\theoremstyle{remark}

\theoremstyle{plain}

\theoremstyle{definitionbreak}
  
\theoremstyle{plainbreak}

\def\LL(#1){\ensuremath{\mathit{LL}(#1)}}
\def\LR(#1){\ensuremath{\mathit{LR}(#1)}}
\def\LALR(#1){\ensuremath{\mathit{LALR}(#1)}}
\def\SLR(#1){SLR(#1)}
\def\SLL(#1){SLL($#1$)}
\def\SLRo(#1){\ensuremath{\mathit{SLR}^\bullet\!(#1)}}

\def\Fi#1(#2){\mathit{First}_{#1}(#2)}
\def\Fo#1(#2){\mathit{Follow}_{#1}(#2)}

\def\FiFo#1(#2,#3){\Fi_{#1}(#3) \cdot_{#1} \Fo_{#1}(#2)}
\def\FIFO#1(#2,#3){\mathit{FiFo}_{#1}(#3,#2)}

\newcommand{\iso}{\cong}

\newcommand{\tf}[1]{\textsf{#1}}

\newcommand{\yields}[1][]{%
  \def\temp{#1}%
  \ifx\temp\empty\operatorname{::=}\else%
    \operatornamewithlimits{::=}\limits_{\tf{#1}}%
  \fi}

\newcommand{\emptyseq}{\varepsilon}        

\newcommand{\rank}{\mathit{rank}}   
\newcommand{\front}{\mathit{front}}               
\newcommand{\rear}{\mathit{rear}}               
\newcommand{\Nat}{\mathbb{N}}
\newcommand{\A}{\mathcal{A}}

\newcommand{\aut}[1]{\mathfrak{#1}}

\renewcommand{\LL}{{\mathbb{L}}}

\makeatletter
\def\tdcfg#1#2{\@ifnextchar[{\@tdcfg{#1}{#2}}{({#1},{#2})}}
\def\@tdcfg#1#2[#3]{({#3},{#1},{#2})} 
\def\cfG(#1)(#2)(#3){%
  \!\left\lceil\frac{#1}{}\!\text{\scriptsize$\circ$}\!\frac{#2}{#3}\right\rceil
}
\makeatother

\newcommand{\CDOT}{{\,\centerdot\,}}

\newcommand{\xyrightarrow}[3]{\mathop{\!\!\!\xymatrix@C=#1{{}\ar@{#2}[r]_{#3}&{}}\!\!\!}\nolimits}
\newcommand{\xyRightarrow}[3]{\mathop{\!\!\!\xymatrix@C=#1{{}\ar@<1pt>@{#2}[r]\ar@<-1pt>@{#2}[r]_{#3}&{}}\!\!\!}\nolimits}
\newcommand{\xyarrow}[2]{%
  \sbox{0}{$\scriptstyle#2$}%
  \mathop{\!\!\!\xymatrix@C\dimexpr\wd0+4pt\relax{{}\ar@{#1}[r]_{#2}&{}}\!\!\!}%
}

\newcommand\Input[1]{\textit{input}({#1})}

\newcommand{\sem}[1]{\llbracket #1 \rrbracket}

\def\buc(#1,#2,#3){[#1] #2 \CDOT #3}

\def\epsilon{\varepsilon}
\def\emptyset{\varnothing}
\def\theta{\vartheta}
\def\rho{\varrho}
\def\phi{\varphi}

\def\autoconf(#1,#2,#3){#3 {\scriptstyle\lozenge} [#1]^{#2}}

\def\pstate(#1,#2){\langle #1, #2 \rangle}

\def\cfaconf(#1,#2){#1 {\scriptstyle\blacklozenge} #2}

\newcommand{\assuref}[1]{\hyperref[#1]{Assumption~\ref*{#1}}}

\newcommand{\detaut}[1]{\aut{#1}_\mathrm{d}}

\newcommand{\sblank}[2]{\emptyseq^{(#1)}_{#2}}

\newcommand{\trystate}{\emph{depthFirst}\xspace}

\SetKwProg{Proc}{procedure}{}{}
\SetKwInOut{Input}{Input}
\SetKwInOut{Output}{Output}
\SetAlgoHangIndent{1em}
\SetNlSty{textsf}{}{}
\SetKwBlock{Either}{either}{}
\SetKwBlock{Or}{or}{}
\SetKw{Continue}{continue}
\SetKw{Call}{call}
\SetKw{Stop}{stop}
\SetFuncSty{upshape}
\SetFuncArgSty{upshape}
\SetProgSty{upshape}

\pgfdeclarelayer{background}              
\pgfdeclarelayer{middleground}
\pgfdeclarelayer{foreground}
\pgfsetlayers{background,middleground,main,foreground}

\newcommand\BackgroundColor{gray!15!white}
\tikzset{
  x=8mm,y=8mm,>=latex,            
  background rectangle/.style
  ={fill=\BackgroundColor,rounded corners=4pt
      },
  glass/.style ={opacity=0,text opacity=0},     
  satin/.style ={opacity=0.3,text opacity=0.3},
  e/.style={inner sep=0.0pt,minimum size=0pt},
o/.style={circle,draw,fill=white,font=\scriptsize,inner sep=1pt,minimum size=2.5mm},
state/.style={circle,draw,fill=white,inner sep=1pt,minimum size=2.5mm},
  }
  {\begin{tikzpicture}%
    [x=6mm,y=-4mm, 
    #1]}%
  {\end{tikzpicture}}
{\begin{tikzpicture}[inner frame sep=4pt,show background rectangle,
    baseline=(current bounding box.center),
    node distance=8mm,
    initial text=,
    every state/.style={fill=white,minimum size=1mm,inner sep=2pt},
    every label/.style={node id},
    #1]}%
  {\end{tikzpicture}}
  {\begin{mygraph}[baseline=(current bounding box.center),#1]}%
  {\end{mygraph}}
\newenvironment{mygraph}[1][x=10mm,y=10mm]%
  {\begin{tikzpicture}[%
      inner frame xsep=0pt,inner frame ysep=4pt,show background rectangle,
      every label/.style={node id},
      #1]}%
    {\end{tikzpicture}}
\def\frontptr(#1){\path (f-#1) edge[-,double distance=1pt,double] (#1)}
\def\rearptr(#1){\path (r-#1) edge[-,double distance=1pt,double] (#1)}
\def\invnode(#1)(#2){
  \node[o,glass] (#1) at (#2) {$#1$};
}
\def\inode(#1)(#2){
  \node[o] (#1) at (#2) {$#1$};
}
\def\fnode(#1)(#2){
  \inode (#1)(#2)
  \node[e] (f-#1) at ($(#2)-(0.5,0)$) {};
  \frontptr(#1);
}
\def\Fnode(#1)(#2)#3{
  \inode (#1)(#2)
  \node[e] (f-#1) at ($(#2)-(#3,0)-(0.5,0)$) {};
  \frontptr(#1);
}
\def\rnode(#1)(#2){
  \inode (#1)(#2)
  \node[e] (r-#1) at ($(#2)+(0.5,0)$) {};
  \rearptr(#1);
}
\def\frnode(#1)(#2){
  \fnode(#1)(#2)
  \node[e] (r-#1) at ($(#2)+(0.5,0)$){};
  \rearptr(#1);
}
  {\begin{tikzpicture}%
     [>=Computer Modern Rightarrow                 
     ,every node/.style={rectangle,inner sep=3pt}
     ,incl/.tip={Hooks[round,right]}               
     ,incL/.tip={Hooks[round,left]}
     ,part/.tip={Circle[open]} 
     ,weak/.tip={Latex[open]} 
     ,harp/.tip={Computer Modern Rightarrow[left]} 
     ,Harp/.tip={Computer Modern Rightarrow[right]}
     ,to/.style={->}
     ,ito/.style={incl->}
     ,ifrom/.style={incL->}
     ,eto/.style={->>}
     ,mto/.style={>->}
     ,pto/.style={part->}
     ,peto/.style={part->>}
     ,pmto/.style={part>->}
     ,tow/.style={-weak}
     ,etow/.style={->weak}
     ,mtow/.style={>-weak}
     ,ptow/.style={part->weak}
     ,petow/.style={part->weak}
     ,pmtow/.style={part>-weak}
     ,x=12mm,y=12mm
     ,#1]}%
  {\end{tikzpicture}}
  {\begin{tikzpicture}[#1]}%
  {\end{tikzpicture}}
\makeatletter

\def\DownEdge[#1]{\tikz \draw (0pt,8pt) edge[#1] (0pt,0pt);}

\def\blob(#1) at (#2)#3{%
  \node[blub](#1)at(#2) {\begin{tabular}{@{}c@{}}  #3 \end{tabular}
};
  }
\def\enlargebb{\node [glass,fit= (current bounding box),inner sep=2pt] {};}

\def\Highlight#1#2#3(#4){
  \begin{pgfonlayer}{background} 
      \node (#4) [subgraph,fit= #1,inner sep=#3,fill=#2] {}; 
   \end{pgfonlayer}
}

\def\Just#1%
  {\tikz \draw (0pt,0pt) edge[#1] (12pt,0pt); 
   \end{graph}}
\def\Graph{\@ifnextchar[{\@Graph}{\@GrapH}}
\def\@Graph[#1]#2{%
  \BOX{%
    \begin{tikzpicture}[x=8mm,y=8mm,label distance=-2pt,>=latex,#1]
     #2%
    \end{tikzpicture}%
  } 
}
\def\@GrapH#1{%
  \BOX{%
    \begin{tikzpicture}[x=8mm,y=8mm,label distance=-2pt,>=latex]
     #1%
    \end{tikzpicture}%
  } 
}

\def\proGraph{\@ifnextchar[{\@proGraph}{\@pro@Graph}}
\def\@proGraph[#1]#2{\BOX{%
  \begin{tikzpicture}[x=8mm,y=-7mm,>=latex,every label/.style={elab},#1]
     #2
    \enlargebb
   \end{tikzpicture}}}
\def\@pro@Graph#1{\BOX{%
  \begin{tikzpicture}[x=8mm,y=-7mm,>=latex,every label/.style={elab}]
    #1
    \enlargebb
  \end{tikzpicture}}}

\def\uvar(#1)#2{\@ifnextchar[{\@uvar(#1)#2}{%
  \node (#1-node) at ($(#1)+(0,1)$) {#2};
  \path[arm] (#1) edge (#1-node);
}}
\def\@uvar(#1)#2[#3]{%
  \node (#1-node) at ($(#1)+(#3,1)$) {#2};
  \path[arm] (#1) edge (#1-node);
}
\def\ustar#1#2{\@ifnextchar[{\@ustar{#1}{#2}}{\Graph{%
      \node (r) [term] at (1,1) {#1};
      \node (h) [nont] at (1,2) {#2};
      \path (r) edge[arm] (h);
}}}
\def\@ustar#1#2[#3]{\Graph[#3]{%
      \node (r) [term] at (1,1) {#1};
      \node (h) [nont] at (1,2) {#2};
      \path (r) edge[arm] (h);
}}

\def\custar#1#2#3{\@ifnextchar[{\@custar(#1)#2}{\proGraph{%
      \node (r) [term] at (1,1) {#1};
      \node (h) [nont] at (1,2) {#2};
      \path (r) edge[arm] (h);
      \node (c) [term] at (1,3) {#3};
}}}
\def\@custar#1#2#3[#4]{\proGraph[#4]{%
      \node (r) [term] at (1,1) {#1};
      \node (h) [nont] at (1,2) {#2};
      \path (r) edge[arm] (h);
      \node (c) [term] at (1,3) {#3};
}}
\makeatother

\newcommand{\BOX}[1]{\begin{array}{@{}c@{}}#1\end{array}}

\title{\GrappaRE\\A Tool for Efficient Graph Recognition Based on\\Finite Automata and Regular Expressions}
\author{Mattia De Rosa
	\institute{Department of Informatics\\
		University of Salerno\\
		Fisciano (SA), Italy}
	\email{matderosa@unisa.it}
	\and
	Mark Minas
	\institute{Computer Science Department\\
		Universit\"at der Bundeswehr M\"unchen\\
		Neubiberg, Germany}
	\email{mark.minas@unibw.de}
}

\newcommand{\titlerunning}{\GrappaRE}
\newcommand{\authorrunning}{M.~De Rosa and M.~Minas}

\hypersetup{
	bookmarksnumbered,
	pdftitle    = {\titlerunning},
	pdfauthor   = {\authorrunning},
	pdfsubject  = {EPTCS},               
}

\begin{document}
\maketitle

\begin{abstract}
  A recent paper by Drewes, Hoffmann, and Minas (GCM 2023 proceedings) has shown
  that certain graph languages can be defined and efficiently recognized by
  finite automata when strings over typed symbols are interpreted as graphs.
  This approach has been implemented in the tool \GrappaRE, which is described
  in this paper. \GrappaRE allows for the convenient specification of graph
  languages through regular expressions, converts each of them into a minimized
  deterministic finite automaton, and checks whether it can recognize graphs
  without the need for backtracking. Measurements confirm that recognition runs
  in linear time.
\end{abstract}

%
\section{Introduction}%
\label{s:intro}

Engelfriet and Vereijken have described that hypergraphs can be composed of
elementary graphs using sequential and parallel composition to define hyperedge
replacement grammars \cite{Engelfriet-Vereijken:97}. If only sequential
composition, i.e., concatenation, is used, then such a composition yields a
string over an alphabet (representing elementary graphs). The resulting graphs
are therefore interpretations of these strings. Finite automata are a common
means of defining regular (string) languages, and by interpreting the generated
strings, such automata define graph languages. This idea is not new, finite
automata for algebraic structures were already studied by Arbib and Give'on in
the 1960s~\cite{Arbib-Giveon:68,Giveon-Arbib:68}, Bozapalidis and Kalampakas
analyzed finite automata on Engelfriet's and Vereijken's
hypergraphs~\cite{Bozapalidis-Kalampakas:08,Kalampakas:11}, Brugging, König
\emph{et al.} in the context of hypergraphs as
cospans~\cite{Blume-Bruggink-Friedrich-Koenig:13}, and more recently Earnshaw
and Soboci\'{n}ski studied regular languages of morphisms in free monoidal
categories, which can be interpreted as hypergraphs or string diagrams, with
their associated automata~\cite{earnshaw_et_al:2022}. All of these approaches
focus on \emph{defining} graph languages through automata.

In contrast, Hoffmann, Drewes, and Minas recently proposed an approach to
efficient graph recognition by finite automata in the sense that an automaton is
``executed'' to decide whether a given graph is a member of its graph language
or not~\cite{drewes-hoffmann-minas:23a}. More precisely, given a finite
automaton and a graph, a decision procedure can decide whether the graph is a
member of the automaton's graph language. This is always possible with an
inefficient backtracking algorithm, but for certain automata backtracking can be
avoided, allowing a recognition in linear time in the number of graph edges.
These automata must satisfy certain conditions, which are presented
in~\cite{drewes-hoffmann-minas:23a} together with algorithms for checking them.

This paper describes the tool \GrappaRE, which implements these checking
algorithms, and the recognition procedure for automata satisfying these
conditions. Special emphasis is placed on the techniques necessary for efficient
graph recognition in linear time. \GrappaRE also supports regular expressions,
which can be used to conveniently specify finite automata. Experiments with this
tool confirm the statement in~\cite{drewes-hoffmann-minas:23a} that it is
possible to recognize graphs with finite automata in linear time. In this way,
\GrappaRE complements the tool landscape of \Grappa\footnote{\Grappa and
	\GrappaRE are available at \url{https://www.unibw.de/inf2/grappa}}. While
\Grappa (\underline{Grap}h \underline{Pa}rsers) is a tool environment for
generating efficient top-down and bottom-up parsers for hyperedge replacement
grammars and contextual hyperedge replacement
grammars~\cite{Drewes-Hoffmann-Minas:15, Hoffmann-Minas:18,
	Drewes-Hoffmann-Minas:19, Drewes-Hoffmann-Minas:21}, \GrappaRE (\Grappa for
\underline{R}egular \underline{E}xpressions) realizes graph recognizers for
regular expressions.

This paper is organized as follows: The next section describes informally the
approach of using finite automata for efficient graph recognition, as proposed
in~\cite{drewes-hoffmann-minas:23a}, and describes in detail the underlying
recognition algorithm with and without backtracking. Then \sectref{s:grappa-re}
describes how this approach has been implemented in \GrappaRE. Special emphasis
is placed on the use of regular expressions as a compact way to specify finite
automata, and the efficient implementation of constant time edge selection,
which is required for graph recognition in linear time. \sectref{s:experiments}
then summarizes the results of experiments with some graph languages specified
by regular expressions and recognizable by \GrappaRE. These results show that
\GrappaRE does indeed recognize graphs in linear time (in the number of their
edges). \sectref{s:concl} concludes the paper.
%
\section{Finite Automata Accepting Graphs}%
\label{s:graphs}
We consider edge-labeled hypergraphs (which we will simply call graphs) without
node labels. A ranked vocabulary is used to label hyperedges (which we will
simply call edges). Each label $\ell$ has a rank $\rank(\ell)\in\Nat$, and each
edge labeled $\ell$ is then attached to $\rank(\ell)$ different nodes. Each
graph $G$ also has a front and a rear interface $\front(G)$ and $\rear(G)$,
respectively, which are repetition-free sequences of nodes in $G$.
Front nodes may also occur in the rear interface. We say that a
graph $G$ is of type $(m,n)$ if $|\front(G)|=m$ and $|\rear(G)|=n$; $m$ is said
to be the front type of $G$ and $n$ its rear type.

\begin{figure}[tb]
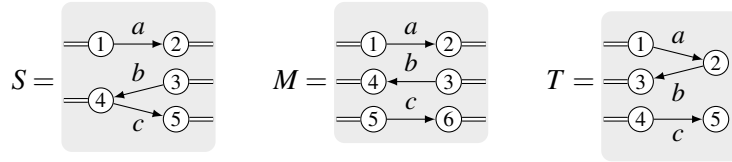

	\centering
	$S=
		\begin{mygraph}[baseline=(3.base),x=10mm,y=-5mm]
			\fnode(1)(1,0)
			\rnode(2)(2,0)
			\rnode(3)(2,1)
			\fnode(4)(1,1.5)
			\rnode(5)(2,2)
			\path
			(1) edge[->] node[above] {\small$a$} (2)
			(3) edge[->] node[above] {\small$b$} (4)
			(4) edge[->] node[below] {\small$c$} (5)
			;
		\end{mygraph}
		\qquad
		M=
		\begin{mygraph}[baseline=(3.base),x=10mm,y=-5mm]
			\fnode(1)(1,0)
			\rnode(2)(2,0)
			\rnode(3)(2,1)
			\fnode(4)(1,1)
			\fnode(5)(1,2)
			\rnode(6)(2,2)
			\path
			(1) edge[->] node[above] {\small$a$} (2)
			(3) edge[->] node[above] {\small$b$} (4)
			(5) edge[->] node[above] {\small$c$} (6)
			;
		\end{mygraph}
		\qquad
		T=
		\begin{mygraph}[baseline=(3.base),x=10mm,y=-5mm]
			\fnode(1)(1,0)
			\inode(2)(2,0.5)
			\fnode(3)(1,1)
			\fnode(4)(1,2)
			\inode(5)(2,2)
			\node at (2.2,0) {};
			\path
			(1) edge[->] node[above] {\small$a$} (2)
			(2) edge[->] node[below] {\small$b$} (3)
			(4) edge[->] node[below] {\small$c$} (5)
			;
		\end{mygraph}$
	\caption{Three graphs $S$, $M$, and $T$.}
	\label{f:graphs}
\end{figure}
\figref{f:graphs} shows three examples of such graphs with the labels $a$, $b$,
and $c$ with $\rank(a)=\rank(b)=\rank(c)=2$. Each graph is contained in a
rectangular box, with circles symbolizing nodes and arrows symbolizing binary
edges. Nodes of the front interface are marked with double lines starting from
the left edge of the box; double lines from the right edge mark nodes of the
rear interface. The order of the double lines from top to bottom defines the
order of the nodes in that interface. For example, the graph $S$ consists of the
nodes 1, 2, 3, 4 and 5 and three edges marked with a, b, and c. The front interface
of $S$ consists of the nodes $1$ and $4$, i.e., the node sequence $1 4$, the
rear interface of the node sequence $2 3 5$. The graphs $S$, $M$, and $T$ are of
types $(2,3)$, $(3,3)$, and $(3,0)$, respectively, because the rear interface of
$T$ is empty.

\begin{figure}[tb]
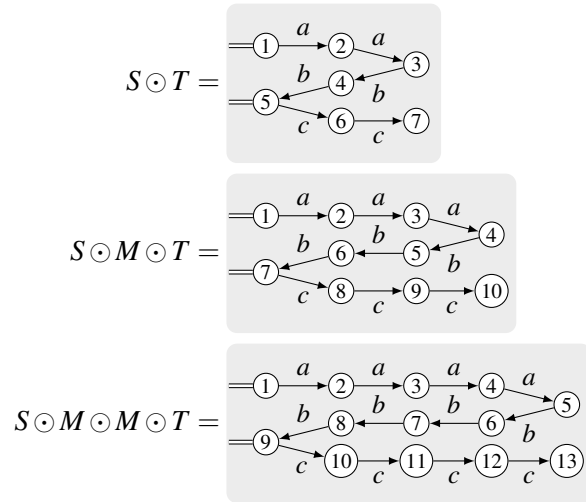

	\begin{align*}
		S \odot T                 & =
		\begin{mygraph}[baseline=(4.base),x=10mm,y=-5mm]
			\fnode(1)(1,0)
			\inode(2)(2,0)
			\inode(3)(3,0.5)
			\inode(4)(2,1)
			\fnode(5)(1,1.5)
			\inode(6)(2,2)
			\inode(7)(3,2)
			\node at (3.2,0) {};
			\path
			(1) edge[->] node[above] {\small$a$} (2)
			(2) edge[->] node[above] {\small$a$} (3)
			(3) edge[->] node[below] {\small$b$} (4)
			(4) edge[->] node[above] {\small$b$} (5)
			(5) edge[->] node[below] {\small$c$} (6)
			(6) edge[->] node[below] {\small$c$} (7)
			;
		\end{mygraph}
		\\
		S \odot M \odot T         & =
		\begin{mygraph}[baseline=(6.base),x=10mm,y=-5mm]
			\fnode(1)(1,0)
			\inode(2)(2,0)
			\inode(3)(3,0)
			\inode(4)(4,0.5)
			\inode(5)(3,1)
			\inode(6)(2,1)
			\fnode(7)(1,1.5)
			\inode(8)(2,2)
			\inode(9)(3,2)
			\inode(10)(4,2)
			\node at (4.2,0) {};
			\path
			(1) edge[->] node[above] {\small$a$} (2)
			(2) edge[->] node[above] {\small$a$} (3)
			(3) edge[->] node[above] {\small$a$} (4)
			(4) edge[->] node[below] {\small$b$} (5)
			(5) edge[->] node[above] {\small$b$} (6)
			(6) edge[->] node[above] {\small$b$} (7)
			(7) edge[->] node[below] {\small$c$} (8)
			(8) edge[->] node[below] {\small$c$} (9)
			(9) edge[->] node[below] {\small$c$} (10)
			;
		\end{mygraph}
		\\
		S \odot M \odot M \odot T & =
		\begin{mygraph}[baseline=(8.base),x=10mm,y=-5mm]
			\fnode(1)(1,0)
			\inode(2)(2,0)
			\inode(3)(3,0)
			\inode(4)(4,0)
			\inode(5)(5,0.5)
			\inode(6)(4,1)
			\inode(7)(3,1)
			\inode(8)(2,1)
			\fnode(9)(1,1.5)
			\inode(10)(2,2)
			\inode(11)(3,2)
			\inode(12)(4,2)
			\inode(13)(5,2)
			\node at (5.2,0) {};
			\path
			(1) edge[->] node[above] {\small$a$} (2)
			(2) edge[->] node[above] {\small$a$} (3)
			(3) edge[->] node[above] {\small$a$} (4)
			(4) edge[->] node[above] {\small$a$} (5)
			(5) edge[->] node[below] {\small$b$} (6)
			(6) edge[->] node[above] {\small$b$} (7)
			(7) edge[->] node[above] {\small$b$} (8)
			(8) edge[->] node[above] {\small$b$} (9)
			(9) edge[->] node[below] {\small$c$} (10)
			(10) edge[->] node[below] {\small$c$} (11)
			(11) edge[->] node[below] {\small$c$} (12)
			(12) edge[->] node[below] {\small$c$} (13)
			;
		\end{mygraph}
	\end{align*}
	\caption{Results of concatenating $S$, $M$, and $T$.}
	\label{f:concatenation}
\end{figure}

Two graphs $G$ and $H$ can be concatenated to form a graph $G \odot H$ if their
types match, which means that the rear type of $G$ is the same as the front type
of $H$.  $G \odot H$ is obtained by taking the disjoint union of $G$ and $H$ and
then merging the corresponding nodes from $\rear(G)$ and $\front(H)$.
\figref{f:concatenation} shows the result of the concatenations $S \odot T$, $S
	\odot M \odot T$, and $S \odot M \odot M \odot T$. If we define $S \odot M^k
	\odot T$ as the graph resulting from the concatenation of $S$, $k$ copies of $M$
and finally $T$, then $S \odot M^k \odot T$ is a graph representation of the
string $a^{k+2}b^{k+2}c^{k+2}$.

The approach in \cite{drewes-hoffmann-minas:23a} is based on the idea of
defining a vocabulary of graph symbols and assigning a specific (elementary)
graph $\sem s$ to each symbol $s$ as an interpretation. Strings over this
vocabulary can then be interpreted as graphs obtained by concatenating the
elementary graphs of the symbols. Of course, this assumes that the string is
valid in the sense that it consists only of concatenations of symbols such that
the corresponding elementary graphs match in type, so that their concatenation
is defined. Otherwise, the string has no valid interpretation.


So-called \emph{blank} and \emph{atom symbols} are defined as vocabulary, with
\emph{blanks} and \emph{atoms} as their graph interpretations. Blanks are
discrete graphs whose front interface contains all nodes of the blank. In
contrast, each atom contains exactly one edge, and no atom has a node that is
neither in its front interface nor attached to its edge. Therefore, all nodes of
the rear interface must also occur in the front interface, be connected to the
edge, or both. These restrictions are necessary for efficient graph recognition,
because in this way every node during the recognition process is either already
contained in the rear interface of the already recognized subgraph, or is
attached to the edge that is read next.

We write blank symbols as $\sblank n\rho$ and atom symbols as $\ell^\phi_\rho$
for any $n\in\Nat$, sequences $\phi$ and $\rho$ over $\{1,\ldots,n\}$ without
repetitions, and edge label $\ell$ with $\rank(\ell)\le n$. The corresponding
blank and atom, $\sem{\sblank n\rho}$ and $\sem{\ell^\phi_\rho}$, respectively,
have nodes $1,\ldots,n$ and $\rho$ as their rear interfaces. $\sem{\sblank
		n\rho}$ has $1\cdots n$, $\sem{\ell^\phi_\rho}$ has $\phi$ as its front
interface. The unique edge of $\sem{\ell^\phi_\rho}$ is labeled $\ell$ and
attached to the nodes $1,\ldots,\rank(\ell)$ (in that order).

For example, \figref{f:interpretation} shows that the graph $S$ in
\figref{f:graphs} is isomorphic to a concatenation of the three atoms
$\sem{a^{13}_{23}}$, $\sem{b^{32}_{312}}$, and $\sem{c^{341}_{342}}$, i.e., $S$
is an interpretation of $a^{13}_{23}\;b^{32}_{312}\;c^{341}_{342}$. Similarly,
$M$ and $T$ are interpretations of $a^{134}_{234}\;b^{324}_{314}\;c^{341}_{342}$
and $a^{134}_{234}\;b^{123}_3\;c^1_\emptyseq$, respectively.
\begin{figure}[tb]
	\centering
	$
		S\quad=\quad
		\begin{mygraph}[baseline=(3.base),x=10mm,y=-5mm]
			\fnode(1)(1,0)
			\rnode(2)(2,0)
			\rnode(3)(2,1)
			\fnode(4)(1,1.5)
			\rnode(5)(2,2)
			\path
			(1) edge[->] node[above] {\small$a$} (2)
			(3) edge[->] node[above] {\small$b$} (4)
			(4) edge[->] node[below] {\small$c$} (5)
			;
		\end{mygraph}
		\quad\iso\quad
		\begin{mygraph}[baseline=(c.base),x=10mm,y=-6mm]
			\fnode(1)(1,0)
			\rnode(2)(2,0)
			\inode(3)(1.5,2)
			\node[e] (f-3) at (0.5,2) {};
			\node[e] (r-3) at (2.5,2) {};
			\frontptr(3);
			\rearptr(3);
			\node[glass,font=\scriptsize] (c) at (1.5,1) {bla};
			\path
			(1) edge[->] node[below] {\small$a$} (2)
			;
		\end{mygraph}
		\odot
		\begin{mygraph}[baseline=(1.base),x=10mm,y=-6mm]
			\rnode(1)(1,1)
			\frnode(2)(1,2)
			\frnode(3)(1,0)
			\path
			(1) edge[->] node[left] {\small$b$} (2)
			;
		\end{mygraph}
		\odot
		\begin{mygraph}[baseline=(4.base),x=10mm,y=-6mm]
			\fnode(1)(1,2)
			\rnode(2)(2,2)
			\inode(3)(1.5,0)
			\node[e] (f-3) at (0.5,0) {};
			\node[e] (r-3) at (2.5,0) {};
			\frontptr(3);
			\rearptr(3);
			\inode(4)(1.5,1)
			\node[e] (f-4) at (0.5,1) {};
			\node[e] (r-4) at (2.5,1) {};
			\frontptr(4);
			\rearptr(4);
			\path
			(1) edge[->] node[below] {\small$c$} (2)
			;
		\end{mygraph}
	$
	\caption{$S$ as in \figref{f:graphs} as an interpretation of
	$a^{13}_{23}\;b^{32}_{312}\;c^{341}_{342}$.}
	\label{f:interpretation}
\end{figure}

A finite automaton over such symbols defines a language of strings accepted by
the automaton, and thus defines a graph language, as long as all these strings
have valid graph interpretations, which is easy to ensure: We define the type of
each symbol to be the type of its elementary graph ($\sblank n\rho$ and
$\ell^\phi_\rho$ then have type $(n,|\rho|)$ and $(|\phi|,|\rho|)$,
respectively), and consider only those automata that allow assigning a rank
$\rank(q)\in\Nat$ to each state $q$, such that each incoming transition of $q$
has a symbol of rear type $rank(q)$, and each outgoing transition of $q$ has a
symbol of front type $\rank(q)$.

\begin{figure}[b]
	\centering
	\begin{tikzpicture}[x=15mm,y=-15mm]
		\node[state] (0) at (0,0) {$q_0$};
		\node[state] (1) at (1,0) {$q_1$};
		\node[state] (2) at (2,0) {$q_2$};
		\node[state] (3) at (2.8,0.5) {$q_4$};
		\node[state] (4) at (2,1) {$q_6$};
		\node[state] (5) at (1,1) {$q_3$};
		\node[state] (6) at (0,1) {$q_5$};
		\node (s) at (-0.6,0) {};
		\node (t) at (-0.6,1) {};
		\path
		(s) edge[->] (0)
		(0) edge[->] node[above] {\small$a^{13}_{23}$} (1)
		(1) edge[->] node[above] {\small$b^{32}_{312}$} (2)
		(1) edge[->] node[left] {\small$b^{12}_2$} (5)
		(2) edge[->,bend left] node[above] {\small$c^{341}_{342}$} (3)
		(3) edge[->,bend left] node[below] {\small$a^{134}_{234}$} (4)
		(4) edge[->] node[left] {\small$b^{324}_{314}$} (2)
		(4) edge[->] node[below] {\small$b^{123}_3$} (5)
		(5) edge[->] node[below] {\small$c^1_\emptyseq$} (6)
		(6) edge[->] (t)
		;
	\end{tikzpicture}
	\caption{Finite automaton accepting graphs in $a^nb^nc^n$.}
	\label{f:abc-auto}
\end{figure}
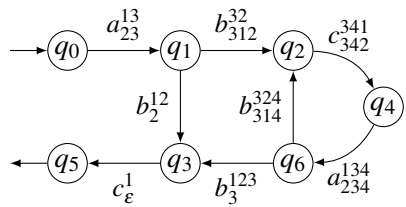

Let $a^nb^nc^n$ denote the language of all graphs representing strings in
$\{a^kb^kc^k \mid k\in\Nat\}$.
Continuing the example above, \figref{f:abc-auto} shows a finite automaton
accepting all strings that can be interpreted as graphs in $a^nb^nc^n$.
$q_0$ is its initial state, $q_5$ its only final
state. The states have the ranks $\rank(q_0)=\rank(q_1)=2$,
$\rank(q_2)=\rank(q_4)=\rank(q_6)=3$, $\rank(q_3)=1$, and $\rank(q_5)=0$.

\subsection{Recognition with Backtracking}
\begin{algorithm}[tb]
	\SetKwFunction{tryState}{depthFirst}%
	\SetKwFunction{tryTransition}{tryTransition}%
	\Input{input graph $G$ and a finite automaton $\aut A$.}
	\Output{\emph{success} if $\aut A$ accepts $G$, and \emph{failure} otherwise.}
	\Begin{
		$q \leftarrow$ initial state of $\aut A$\;
		$F\leftarrow$ front interface of $G$\;
		$U\leftarrow$ all edges of $G$\;
		\lIf{all discrete nodes of $G$ occur in $F$\label{r:discrete}}{
			\Call \tryState{$q,F,U$}%
		}
		\Stop with \emph{failure}\label{r:fail-a}
	}
	\BlankLine
	\Proc{\tryState{$q$: State, $F$: Sequence of nodes, $U$: Set of edges}}{
		\lIf{$q$ is final, $F$ is the rear interface of $G$, and $U=\emptyset$}{
			\Stop with \emph{success}\label{r:terminate}%
		}
		\ForEach{outgoing transition $t$ of $q$\label{r:all-trans}}{
			let $q'$ be the state reached by $t$\;
			\If{$t$ is an atom transition}{
				let $\alpha$ be the atom symbol of $t$\;
				\ForEach{edge $e\in U$ as specified by $\alpha$ and
					attached to nodes in $F$\label{l:select-e}}{
					let $F'$ be $F$ after modifying it according to
					$\alpha$ and $e$\;
					\Call \tryState{$q',F',U \setminus \{e\}$}
				}
			}
			\Else {
				let $F'$ be $F$ after modifying it according to
				the blank symbol of $t$\label{r:new-blank-F}\;
				\Call \tryState{$q',F',U$}\label{r:rec-try-state}
			}
		}
	}
	\caption{Recognizing a graph with a finite automaton using backtracking.}
	\label{alg:recognize}
\end{algorithm}
While interpreting a string accepted by an automaton is straightforward,
recognizing a given graph, i.e., finding a string representation of the graph
such that the string is accepted by the automaton, involves a search process,
usually with backtracking due to nondeterministic decisions.
\algref{alg:recognize} outlines this process as a classic depth-first search
with backtracking using the recursive \trystate procedure. It tries to find a
walk through the finite automaton from the initial state to some final state
that accepts a string with the input graph $G$ as its interpretation. It either
fails in \alineref{r:fail-a} or terminates successfully in
\alineref{r:terminate}.

The algorithm does not call the \trystate procedure but fails immediately if $G$
has a discrete node that is not in its front interface (\alineref{r:discrete}
and \alineref{r:fail-a}). Recall that no atom or blank can create a discrete
node, since all nodes in the rear interface also occur in the front interface or
are attached to the (unique) edge of the atom.

The \trystate procedure is called with three parameters $q,F,U$: $q$ is a state
of the automaton, $F$ (for ``front'') contains a sequence of nodes that is the
front interface of the next elementary graph to be read, and $U$ (for
``unread'') contains all edges of $G$ that have not yet been read. The procedure
finally terminates successfully if it can (recursively) find a walk from $q$ to
some final state such that it reads all edges in $U$, starting with $F$ as the
front interface and ending with the rear interface of $G$. This means that if
\trystate is called with $q$ being a final state, $F$ being the rear interface
of $G$, and $U=\emptyset$, \algref{alg:recognize} will terminate successfully
(\alineref{r:terminate}). Otherwise \trystate tries each of the outgoing
transitions of $q$ in the usual depth-first manner in \alineref{r:all-trans},
i.e., it selects an arbitrary transition first and continues the search with
recursive procedure calls to \trystate (see below). These calls return only if
they cannot find a successful walk. The next transition is then selected in
\alineref{r:all-trans}, and so on. The current \trystate call is terminated,
i.e., \algref{alg:recognize} backtracks, if all outgoing transitions of $q$ lead
to a dead end.

If the selected transition in \alineref{r:all-trans} is an atom transition, i.e,
if it is labeled with an atom symbol $\alpha$, it tries to find a matching edge
$e$ not yet read (\alineref{l:select-e}); $\alpha$ and $F$ determine which edge
may be selected: $\alpha$ determines the label of $e$ and how it must be
connected to nodes in $F$. In particular, the front nodes of $\sem \alpha$ must
correspond to the nodes in $F$. Nodes that are not in $F$ but are attached to
$e$ must be ``new'' nodes, i.e., they must not have been encountered before. The
latter follows from the fact that nodes of $\sem \alpha$ that do not appear in
its front interface are created by this step. If the recognizer has several
candidates to choose from in \alineref{l:select-e}, it will select them one by
one by recursively calling \trystate. If this call fails, it will try the next
one, and so on.

However, if the selected transition in \alineref{r:all-trans} is labeled with a
blank symbol, no further decision is required; a blank can always be processed,
it just modifies the rear interface of the subgraph of $G$ read so far
(\alineref{r:new-blank-F}) and continues the depth-first search by recursively
calling \trystate in \alineref{r:rec-try-state}.\footnote{Note that a cycle of
	blank transitions can lead to an infinite recursion, which could easily be
	prevented by keeping track of the transitions visited so far. However, this has
	been omitted here to simplify the presentation.}

\subsection{Recognition without Backtracking}

This search can be expensive because it uses backtracking to try out possible
solutions instead of systematically selecting suitable transitions and edges,
which can lead to exponential runtimes. To address this problem,
\cite{drewes-hoffmann-minas:23a} has discussed conditions under which
backtracking can be avoided completely in this search process. In fact, the
automaton must satisfy additional conditions to allow a deterministic
recognition process without backtracking: It must have the \emph{transition
	selection} (TS) property and the \emph{free edge choice} (FEC) property, which
will be explained below. Since backtracking is then not necessary,
\algref{alg:recognize} is transformed into a simplified version
(\algref{alg:efficiently-recognize}), which will be discussed next.

It is clear that a deterministic choice of a suitable transition is impossible
if the automaton~$\aut A$ is nondeterministic. That is, if it contains a state
with two outgoing transitions labeled with the same graph symbol. Therefore, the
finite automaton must first be transformed into a deterministic finite automaton
(DFA), as in the case of transforming a nondeterministic finite automaton for
strings, before it can be used by \algref{alg:efficiently-recognize}. The
approach proposed in \cite{drewes-hoffmann-minas:23a} uses a modified version of
the well-known powerset construction. Surprisingly, however, the powerset
construction does not always produce a deterministic automaton in this context.
This can happen if $\aut A$ uses atom symbols which differ only in their rear
interfaces (see \cite[Example~4.1]{drewes-hoffmann-minas:23a}). In this case,
however, one can always transform $\aut A$ into an equivalent automaton $\aut
	A'$, which can then be transformed into an equivalent DFA $\detaut A$ using the
modified powerset construction. The transformation of $\aut A$ into the DFA
$\detaut A$ is thus a two-step process, consisting of the preprocessing step (called
\emph{disambiguation}) and the following (modified) powerset construction.

\begin{algorithm}[tb]
	\caption{Recognizing a graph with a DFA that has the FEC and TS property.}
	\label{alg:efficiently-recognize}
	\Input{input graph $G$ and a DFA $\detaut A$ that has the FEC and TS property.}
	\Output{\emph{success} if $\detaut A$ accepts $G$, and \emph{failure} otherwise.}
	\lIf{$G$ contains discrete nodes that are not in its front interface}{%
		\Stop with \emph{failure}%
	}
	$q\leftarrow$ initial state of $\detaut A$\;
	$F\leftarrow$ front interface of $G$\;
	$U\leftarrow$ all edges of $G$\;
	\While{$U\neq\emptyset$ \label{b:begin-while}}{
		\ForEach{outgoing atom transition $t$ of $q$\label{b:begin-foreach}}{
			let $\alpha$ be the atom symbol of $t$\;
			\If{there is an edge $e\in U$ as specified by $\alpha$ and attached
				to nodes in $F$}{
				let $e$ be such an edge\label{b:select-edge}\;
				modify $F$ according to $\alpha$ and $e$\label{b:new-F}\;
				set $q$ to the state reached by $t$\;
				remove $e$ from $U$\;
				\Continue while loop at \alineref{b:begin-while}\label{b:end-foreach}%
			}
		}
		\Stop with \emph{failure}\label{b:end-while}
	}
	\lIf{$q$ is a final state and $F$ is the rear interface of $G$}{
		\Stop with \emph{success}\label{b:success-a}%
	}
	\lIf{there is an outgoing blank transition of $q$ such that its blank turns $F$ into the rear interface of $G$}{
		\Stop with \emph{success}\label{b:success-b}%
	}
	\Stop with \emph{failure}
\end{algorithm}
If $\detaut A$ produced by the modified powerset construction contains blank
transitions, all of them reach a final state that has no outgoing transitions.
Consequently, a blank transition can only be selected if all edges of $G$ are
have been read. All previously selected transitions must be atom transitions.
Each loop cycle of the while loop (\alinesref{b:begin-while}--\ref{b:end-while}
of \algref{alg:efficiently-recognize}) selects an atom transition and then reads
a corresponding edge of $G$. The TS property of $\detaut A$ implies that the
only possible transition can be selected as shown in
\algref{alg:efficiently-recognize}: The foreach loop
(\alinesref{b:begin-foreach}--\ref{b:end-foreach}) tries all outgoing atom
transitions of the current state in a predefined order until it finds the first
one such that $G$ contains a matching edge that has not yet been read before.
This order is determined by the algorithm that checks the TS property of
$\detaut A$ described in \cite{drewes-hoffmann-minas:23a}. In fact, $\detaut A$
has the TS property if such an order exists for each of its states.

Note that \algref{alg:efficiently-recognize} in \alineref{b:select-edge} selects
any unread edge $e$ that matches the atom symbol $\alpha$ of the selected
transition. In fact, every edge that matches $\alpha$ is equally suitable if
$\detaut A$ has the FEC property: If one of these candidates is the right choice
for $\detaut A$ to accept $G$, then all of them are.
\cite{drewes-hoffmann-minas:23a} has proposed a sufficient criterion for
checking the FEC property, i.e., $\detaut A$ has the FEC property if the
criterion is met. But if the criterion is not met, $\detaut A$ may or may not
have the FEC property. The criterion consists of identifying all transitions in
$\detaut A$ that are \emph{deferrable}. These are transitions where
\algref{alg:efficiently-recognize} can choose between several possible
candidates in \alineref{b:select-edge}. They are called deferrable because one
of the candidates is selected now, while the others must be selected in later
cycles of the while loop, possibly using other transitions.
\cite{drewes-hoffmann-minas:23a} has outlined how to identify deferrable
transitions. The criterion now requires that the rear interface of $\alpha$ does
not contain any nodes that are not also contained in its front interface, i.e.,
the selection of $e$ in \alineref{b:select-edge} cannot affect the selection of
$F$ in \alineref{b:new-F}.

\algref{alg:efficiently-recognize} terminates successfully after it has read all
edges of $G$ in its while loop and has either reached a final state where $F$
and the rear interface of $G$ are identical (\alineref{b:success-a}), or it can
reach a final state by taking a blank transition that modifies $F$ in the rear
interface of $G$ (\alineref{b:success-b}). Otherwise, $\aut A$ does not accept
$G$ and \algref{alg:efficiently-recognize} terminates with a failure.

As noted above, a single edge is read in each cycle of the while loop in
\alinesref{b:begin-while}--\ref{b:end-while}.
\cite{drewes-hoffmann-minas:23a}~argued that a loop cycle takes only constant
time if edge selection is implemented properly. Consequently,
\algref{alg:efficiently-recognize} takes only linear time (in the number of
edges) to recognize an input graph. This is demonstrated in
\sectref{s:experiments}.

%
\section{\GrappaRE}%
\label{s:grappa-re}
\begin{figure}[tbp]
	\centering
	\begin{tikzpicture}[font={\fontsize{7pt}{7}\selectfont},x=20.8mm,y=-13mm]
		\sffamily
		\node at (0,1) [event,label={[align=center]Use automaton\\spec}] (start-automaton) {};
		\node at (0,3) [event,node distance=18.5mm,label={[align=center]Use regular\\expression}] (start-re) {};
		\node at (1,3) [task,align=center] (re-reader) {RegExp\\Reader};
		\node at (1,4) [data object,style={fill=yellow},align=center] (re-spec) {RegExp\phantom{-}\\spec};
		\node at (1,1) [task,align=center] (automaton-reader) {Automaton\\Reader};

		\node at (1,0) [data object,style={fill=yellow},align=center] (automaton-spec) {Automaton\phantom{-}\\spec};
		\node at (1,2) [data object] (aut-A) {$\aut A$};
		\node at (2,3) [task,align=center] (re-to-automaton) {RegExp\\Translator};
		\node at (2,4) [data object] (re) {RegExp};

		\node at (2,1) [task,align=center] (disambiguation) {Disam-\\biguation};
		\node at (3,0) [data object] (aut-Ap) {$\aut A'$};

		\node at (3,1) [task,align=center] (powerset-construction) {Powerset\\Construction};

		\node at (4,1) [task,align=center] (automaton-minimizer) {Automaton\\Minimizer};
		\node at (4,0) [data object] (detaut-A) {$\detaut A$};

		\node at (5,1) [task,align=center] (fec-check) {FEC\\Check};
		\node at (5,2) [data object] (detaut-Ap) {$\detaut A'$};

		\node at (6,1) [exclusive gateway] (gateway1) {};
		\node at (7,1) [end event] (end-fail1) {};

		\node at (6,2) [task,align=center] (ts-check) {TS\\Check};

		\node at (7,3) [end event] (end-fail2) {};
		\node at (6,3) [exclusive gateway] (gateway2) {};

		\node at (5,3) [task,align=center] (reorder-transitions) {Reorder\\Transitions};
		\node at (5,4) [data object] (detaut-App) {$\detaut A''$};

		\node at (4,3) [task] (recognizer) {Recognizer};
		\node at (4,2) [data object,style={fill=yellow}] (graph) {Graph};
		\node at (4,4) [data object,style={fill=yellow}] (results) {Results};
		\node at (3,3) [end event] (end-success) {};

		\draw[sequence,->] (start-re) -- (re-reader);
		\draw[message,->] (re-spec) -- (re-reader);
		\draw[sequence,->] (start-automaton) -- (automaton-reader);
		\draw[message,->] (automaton-spec) -- (automaton-reader);
		\draw[message,->] (automaton-reader) -- (aut-A);
		\draw[sequence,->] (re-reader) -- (re-to-automaton);
		\draw[message,->] (re-to-automaton) -- (aut-A);
		\draw[message,->] (re-reader) -- (re);
		\draw[message,->] (re) -- (re-to-automaton);

		\draw[sequence,->] (automaton-reader) -- (disambiguation);
		\draw[message,->] (aut-A) -- (disambiguation);
		\draw[sequence,->] (re-to-automaton) -- (disambiguation);
		\draw[message,->] (disambiguation) -- (aut-Ap);

		\draw[sequence,->] (disambiguation) -- (powerset-construction);
		\draw[message,->] (aut-Ap) -- (powerset-construction);

		\draw[message,->] (powerset-construction) -- (detaut-A);
		\draw[sequence,->] (powerset-construction) -- (automaton-minimizer);
		\draw[message,->] (detaut-A) -- (automaton-minimizer);

		\draw[message,->] (automaton-minimizer) -- (detaut-Ap);
		\draw[sequence,->] (automaton-minimizer) -- (fec-check);
		\draw[message,->] (detaut-Ap) -- (fec-check);

		\draw[sequence,->] (fec-check) -- (gateway1);
		\draw[sequence,->] (gateway1) -- (end-fail1) node [pos=0.4,above] {Fail};

		\draw[message,->] (detaut-Ap) -- (ts-check);
		\draw[sequence,->] (gateway1) -- (ts-check) node [pos=0.6,above] {Success};

		\draw[sequence,->] (ts-check) -- (gateway2);
		\draw[sequence,->] (gateway2) -- (end-fail2) node [pos=0.4,above] {Fail};

		\draw[message,->] (detaut-Ap) -- (reorder-transitions);
		\draw[sequence,->] (gateway2) -- (reorder-transitions) node [pos=0.4,above] {Success};
		\draw[message,->] (reorder-transitions) -- (detaut-App);

		\draw[message,->] (detaut-App) -- (recognizer);
		\draw[sequence,->] (reorder-transitions) -- (recognizer);
		\draw[message,<-] (recognizer) -- (graph);
		\draw[message,->] (recognizer) -- (results);
		\draw[sequence,->] (recognizer) -- (end-success);
	\end{tikzpicture}
	\caption{Architecture of \GrappaRE}
	\label{f:arch}
\end{figure}

The approach described in~\cite{drewes-hoffmann-minas:23a} and summarized in the
previous section has been realized in the tool \GrappaRE. It is implemented in
\Java and uses several external packages, in particular \Brics~\cite{Brics} with
its standard algorithms for finite automata and regular expressions for strings,
and \Graphstream~\cite{GraphStream} for graph visualization. The architecture of
\GrappaRE is represented as a BPMN\footnote{Business Process Management
	Notation} diagram in \figref{f:arch}. As usual, rectangles represent activities,
document icons represent data, either external files (e.g., \emph{Automaton
	spec}) read or written by \GrappaRE, which have a yellow background, or internal
data (e.g., $\aut A$), drawn with a white background. The gateways represent
decisions where only one of the outgoing paths is taken.

\GrappaRE can be used in two modes, represented by the two different start
events, i.e., either with the specification of a finite automaton (``\emph{Use
	automaton spec}''), or with a regular expression (``\emph{Use regular
	expression}''). We will describe the first mode first, then \sectref{s:regexp}
will describe the second mode. Both modes can be used interactively with a GUI
or in a headless variant. The GUI is described in \sectref{s:GUI}, and the
headless variant was used in the experiments to check the speed of the
recognition procedure described in \sectref{s:experiments}.

\GrappaRE reads finite automata and input graphs from text files. In the first
mode (``\emph{Use automaton spec}''), the \emph{Automaton Reader} reads the
specification of a finite automaton from a text file. \figref{f:abc-auto-spec}
shows such a textual specification of the automaton in \figref{f:abc-auto}.
\begin{figure}[b]
	\centering
	\fbox{
		\small
		\begin{minipage}{13.7cm}
\begin{verbatim}
auto abc {
  symbol a(2), b(2), c(2);
  state q0(2), q1(2), q2(3), q3(1), q4(3), q5(0)*, q6(3);
  start q0;
  q0 -- a^13_23   --> q1;  q1 -- b^32_312 --> q2;  q1 -- b^12_2    --> q3;
  q2 -- c^341_342 --> q4;  q3 -- c^1_<>   --> q5;  q4 -- a^134_234 --> q6;
  q6 -- b^324_314 --> q2;  q6 -- b^123_3  --> q3;
}
\end{verbatim}
		\end{minipage}
	}
	\caption{\GrappaRE specification of the automaton in \figref{f:abc-auto}.}
	\label{f:abc-auto-spec}
\end{figure}
The specification must contain the ranked alphabet of edge labels after the
\texttt{symbol} keyword; the rank of each symbol is given in parentheses. The
states are listed after the \texttt{state} keyword. State ranks (see
\sectref{s:graphs}) are again given in parentheses. Final states are marked
with~\texttt{*}, the start state with the \texttt{start} keyword, and
transitions in the the obvious way. For example, \verb|a^13_23| represents the
atom symbol $a^{13}_{23}$ and \verb|c^1_<>| represents the atom symbol
$c^1_\emptyseq$. Blank symbols such as $\sblank{2}{12}$ are written as
\verb|<>^2_12|, but do not appear in \figref{f:abc-auto-spec}.

The \emph{Automaton Reader} produces an internal representation $\aut A$ of the
automaton and then transforms it into a DFA in the two-step process outlined in
\sectref{s:graphs}: \emph{Disambiguation} is the preprocessing step that ensures
that the \emph{Powerset Construction} really produces a deterministic finite
automaton, denoted here by $\detaut A$.

Similar to the string case, the DFA $\detaut A$ can contain nondistinguishable
states. These are states that accept the same languages over graph symbols when
starting in these states. Since sequences of graph symbols are special cases of
strings, one can use any standard algorithm to minimize $\detaut A$ by merging
all non-distinguishable states. \GrappaRE uses Hopcroft's
algorithm~\cite{Hopcroft:1971} with an implementation provided by
\Brics~\cite{Brics}, which yields an equivalent DFA $\detaut A'$.

The following steps in \figref{f:arch} check if backtracking can be avoided when
using $\detaut A'$ to recognize graphs. \emph{FEC Check} checks if it has the
FEC property (see \sectref{s:graphs}), and \emph{TS Check} checks if it has the
TS property using the algorithms described in~\cite{drewes-hoffmann-minas:23a}.
If one of them fails, efficient graph recognition cannot be guaranteed, and
\GrappaRE refuses to use $\detaut A'$ for graph recognition. However, if both
succeed, \emph{Reorder Transitions} then reorders the transitions of each state
as required by \algref{alg:efficiently-recognize} and described in
\cite{drewes-hoffmann-minas:23a}. In fact, this is done by topologically sorting
the (atom) transitions of each state according to a partial order obtained by
the TS check. The output of this step is a DFA~$\detaut A''$.

$\detaut A''$ is finally used in the \emph{Recognizer} step that implements
\algref{alg:efficiently-recognize}. The input graph $G$ is either read from an
external file in \Json format, or it can be generated programmatically in
headless mode. This step checks if $\detaut A''$ (and therefore $\aut A$)
accepts $G$ by finding a sequence $s$ of graph symbols such that $G\iso\sem s$.
The result of this step, which can be visually inspected in GUI mode, is
described in \sectref{s:GUI}.

The \emph{Recognizer} step tries to recognize the input graph as described by
\algref{alg:efficiently-recognize}, i.e., by selecting suitable edges of $G$
step by step. \GrappaRE provides two different implementations of this edge
selection: a simple implementation searches for a suitable edge iteratively and
thus takes linear time (in the number of edges of $G$). Graph recognition then
takes quadratic time, as shown in \sectref{s:experiments}. However, \GrappaRE
also provides a more efficient selection implementation, described in
\sectref{sec:EfficientEdgeSelection}, which takes only constant time. Thus,
\GrappaRE can recognize graphs in linear time, as claimed
in~\cite{drewes-hoffmann-minas:23a} and demonstrated in \sectref{s:experiments}.

\subsection{Regular Expressions}%
\label{s:regexp}

\GrappaRE also provides a second mode where the finite automaton is not
specified directly as in \figref{f:abc-auto-spec}, but by a regular expression.
These regular expressions use graph symbols as elementary symbols, but are
otherwise structured like normal regular expressions. In particular, they use
concatenation and alternatives, as well as the Kleene star, with the same
semantics as in the string case. Each such expression corresponds in the usual
way to a finite automaton with the same language. However, not every regular
expression over graph symbols is a valid expression. Rather, the finite
automaton corresponding to the expression may only accept valid graph
interpretations as sequences of graph symbols, i.e., the types of successive
graph symbols must match. Just as this can be ensured for automata (see
\sectref{s:graphs}), it is also possible for regular expressions, as long as it
is not just $\emptyseq$, because it does not have any type information, i.e., how
many front and rear nodes a corresponding graph has. Consequently, $\emptyseq$
is not a valid regular expression.

For example, the automaton shown in \figref{f:abc-auto} and
\figref{f:abc-auto-spec} can be more compactly described by the regular expression
$$
	a^{13}_{23} \; b^{12}_2 \; c^1_\emptyseq
	\;\;\Big|\;\;
	a^{13}_{23} \; b^{32}_{312} \; c^{341}_{342} \;
	(a^{134}_{234} \; b^{324}_{314} \; c^{341}_{342})^* \;
	a^{134}_{234} \; b^{123}_3 c^1_\emptyseq.
$$

\begin{figure}[tb]
	\centering
	\fbox{\small
		\begin{minipage}{7cm}
\begin{verbatim}
regexp abc {
  symbol a(2), b(2), c(2);

  a^13_23 b^12_2 c^1_<>
  | 
  a^13_23 b^32_312 c^341_342 
  (a^134_234 b^324_314 c^341_342)* 
  a^134_234 b^123_3 c^1_<>
} 
\end{verbatim}
		\end{minipage}
	}
	\caption{Regular expression for the automaton in \figref{f:abc-auto}.}
	\label{f:abc-regexp-spec}
\end{figure}
Step  \emph{RegExp Reader} (see \figref{f:arch}) can read such a regular
expression in textual format as in \figref{f:abc-regexp-spec}. This
specification must also include a declaration of the ranked vocabulary of edge
labels. Graph symbols are then specified with the same syntax as in automata
specifications (e.g., \figref{f:abc-auto-spec}).

\emph{RegExp Reader} also checks if the regular expression is valid, and
\emph{RegExp Translator} then transforms it into an equivalent finite automaton,
yielding $\aut A$ as an internal representation of the resulting finite
automaton. To implement this step, it again uses \Brics~\cite{Brics}, which
provides an implementation for the string case that can be used directly here.
The automaton $\aut A$ is then used in the same way as in the mode ``\emph{Use
	automaton spec}''.

\subsection{Efficient Edge Selection}
\label{sec:EfficientEdgeSelection}

As described above, \GrappaRE implements edge selection as used in
\alineref{b:select-edge} of \algref{alg:efficiently-recognize} in a
straightforward simple way and in an efficient proper way. We will refer to them
as the \emph{simple} and \emph{efficient} implementations, respectively.

To speed up edge selection, both implementations build a data structure before
starting the recognition process. The simple implementation simply collects
lists of edges, where each list contains all edges with the same label. When
\alineref{b:select-edge} needs to select an edge as specified by an atom symbol
$\ell^\phi_\rho$, it inspects the list for~$\ell$ and looks for the first edge
attached to the current nodes in~$F$ as specified by~$\phi$. Once a matching
edge is found, it can be efficiently removed from the list because it is
implemented as a doubly linked list. However, the sequential search for this
edge takes linear time in the number of edges with the label~$\ell$.
\sectref{s:experiments} shows that graph recognition with this simple
implementation then takes quadratic time and is rather slow.

The efficient implementation builds more sophisticated data structures before
starting the recognition process and uses techniques similar to those in
\Grappa for efficient graph parsing~\cite{Hoffmann-Minas:18}. It uses a
two-step process:

\begin{enumerate}
	\item In the first step, the automaton $\detaut A''$ (see \figref{f:arch})
	      is analyzed and all atom symbols in $\detaut A''$ are collected in a
	      set. This set contains all atom symbols that can specify an edge
	      selection (as in \alineref{b:select-edge} in
	      \algref{alg:efficiently-recognize}). Each of the atom symbols
	      $\ell^\phi_\rho$ specifies how the edge to be selected must be
	      connected to the current nodes in~$F$. The other nodes of the edge,
	      i.e., those that are not referred to by~$\phi$, must be nodes that have
	      not been encountered before. This information is then used in the
	      second step to collect suitable lists of edges, which speeds up edge
	      selection.
	\item In the second step, lists of edges of the input graph $G$ are
	      collected, one list for each situation that can occur during edge
	      selection (as in \alineref{b:select-edge} in
	      \algref{alg:efficiently-recognize}). Each of these lists will contain
	      all edges that are suitable in the corresponding situation.
\end{enumerate}

As an example, consider the graph language $a^nb^nc^n$ and its DFA in
\figref{f:abc-auto}. It contains the atom symbols $a^{13}_{23}$,
$a^{134}_{234}$, $b^{12}_2$, $b^{123}_3$, $b^{32}_{312}$, $b^{324}_{314}$,
$c^1_\emptyseq $, and $c^{341}_{342}$. Now consider the situation of selecting
an edge that matches $\ell^\phi_\rho=a^{13}_{23}$, $F$ in
\algref{alg:efficiently-recognize} must consist of two nodes, say $F=(m_1,m_2)$.
$1$ at position $1$ of $\phi=13$ indicates that one must choose an $a$-edge
attached to two nodes $(n_1,n_2)$ such that $n_1=m_1$, and $n_2$ is not yet read
since $2$ does not occur in $\phi$. Finding such an edge is easy and takes only
constant time if each node has a list of all $a$-edges connected to that node as
its first node. This is the usual association list approach. Note, however, that
one cannot select an $a$-edge whose second node has already been read.
Consequently, one must remove all $a$-edges from the corresponding association
lists as soon as their second attached node has been encountered. This can also
be done efficiently if each edge keeps the information in which association
lists it is stored. Their number has an upper bound, which can be read from the
set of atom symbols of $\detaut A''$. And removing edges from all these lists
can also be done efficiently if these lists are doubly linked lists and one
keeps references to the corresponding list buckets. Therefore, removing an edge
from all corresponding lists also takes constant time.

\begin{figure}[b]
	\includegraphics[scale=0.27]{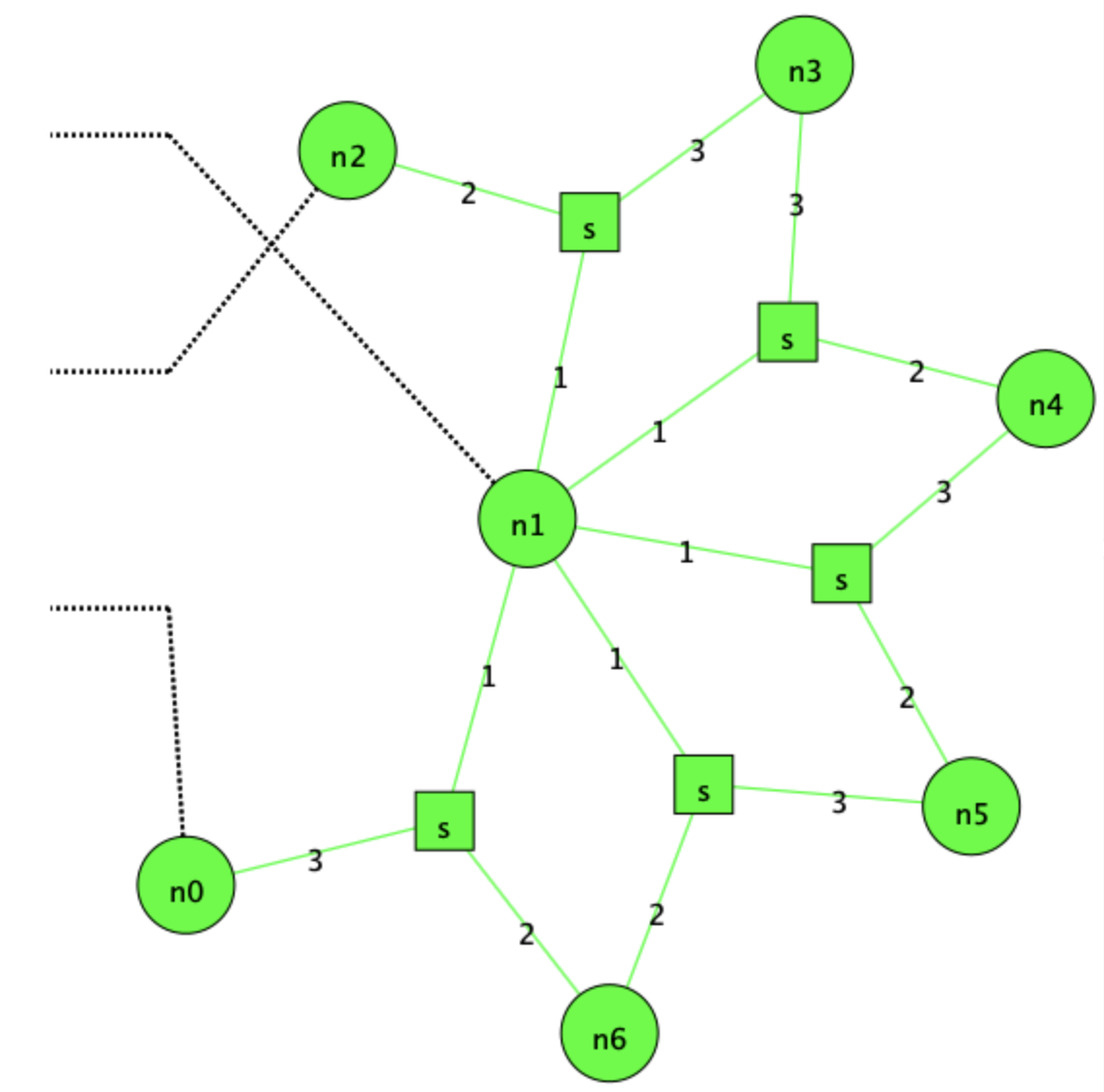}
	\hfill
	\includegraphics[scale=0.27]{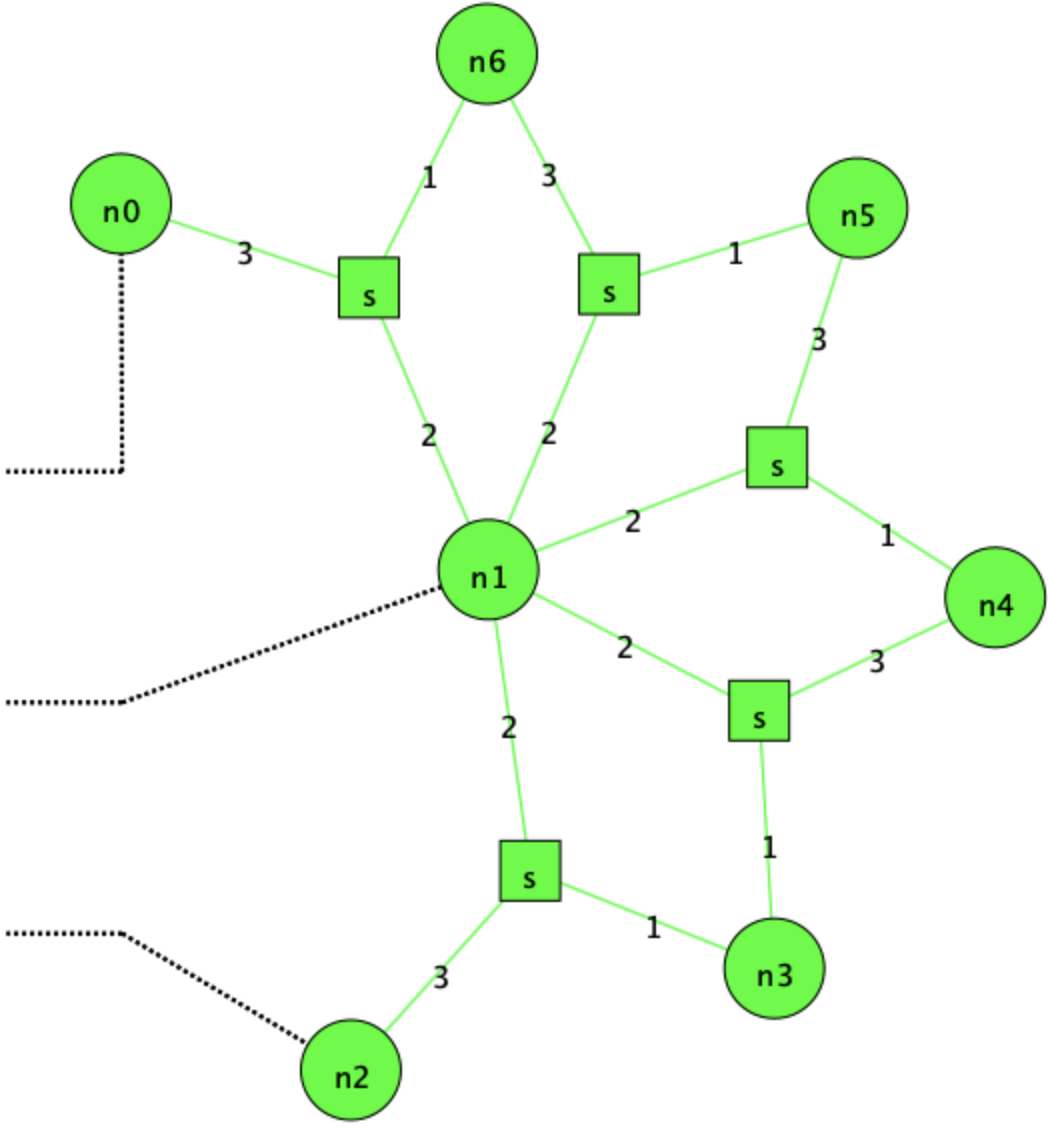}
	\hfill
	\includegraphics[scale=0.27]{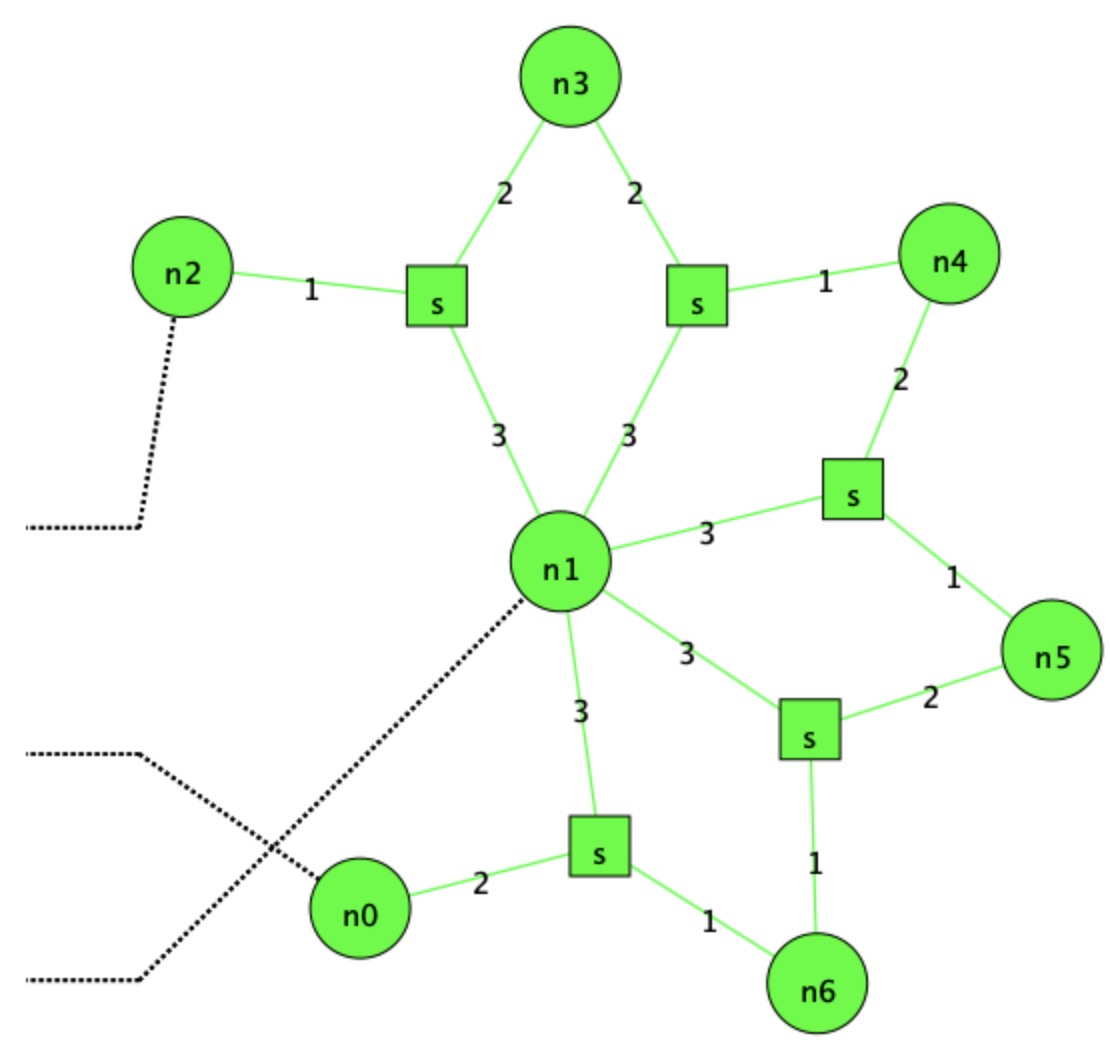}
	\caption{Three \Spikes graphs.}
	\label{f:suns}
\end{figure}
However, keeping only simple association lists is not always sufficient to ensure
constant time edge selection. To see this, consider the \Spikes graph language
represented by
$$
	s^{124}_{134} \; (s^{134}_{124})^\ast \; s^{123}_\emptyseq
	\;\;\Big|\;\;
	s^{423}_{421} \; (s^{421}_{423})^\ast \; s^{321}_\emptyseq
	\;\;\Big|\;\;
	s^{143}_{243} \; (s^{243}_{143})^\ast \; s^{123}_\emptyseq
$$%
where label $s$ has rank 3. We call these graphs \Spikes because of their shape;
\figref{f:suns} shows three of them. Nodes are drawn as circles, $s$-edges as
rectangles connected to their attached nodes. Numbers indicate the order of the
attachments. \Spikes graphs are of type $(3,0)$. Nodes of their front interfaces
are indicated by lines to the left border, ordered from top to bottom. Graph
recognition must start with a unique edge; for the graphs in \figref{f:suns}
this is always the edge attached to $n_1,n_2,n_3$. The central node $n_1$, which
is connected to all edges of the graph, can be any node of its front interface,
as shown in \figref{f:suns}. Consequently, any front node can be attached to any
number of edges. Thus, it is impossible to select the unique starting edge in
constant time by simply accessing an association list of a front node. Instead,
one must be able to look up edges by pairs of their nodes, as can be seen in the
following example: The first alternative of the regular expression starts with
$s^{124}_{134}$, i.e., we are looking for an $s$-edge whose first and second
attached nodes are the first and second nodes in its front interface, i.e.,
$n_1$ and $n_2$ for the left \Spikes graph. This is only the case for the edge
attached to $n_1,n_2,n_3$.

\GrappaRE uses hash tables to manage association lists for tuples of nodes and
to allow efficient edge lookup. The association lists are created and filled
with edges before recognition starts. The hash table lookup, and thus the
association list lookup, takes constant time on average because each hash table
is fixed after preprocessing; only the contents of the association lists are
modified in the same way as described above. Since the number of atom symbols
occurring in a DFA is fixed, all these data structures require linear space in
the size of the input graph if each hash table size is chosen to be proportional
to the number of all edges. In addition, on average, setting up these data
structures takes linear time in the size of the input graph.

\subsection{Graphical User Interface}%
\label{s:GUI}%
\begin{figure}[tb]
	\includegraphics[width=.7\textwidth]{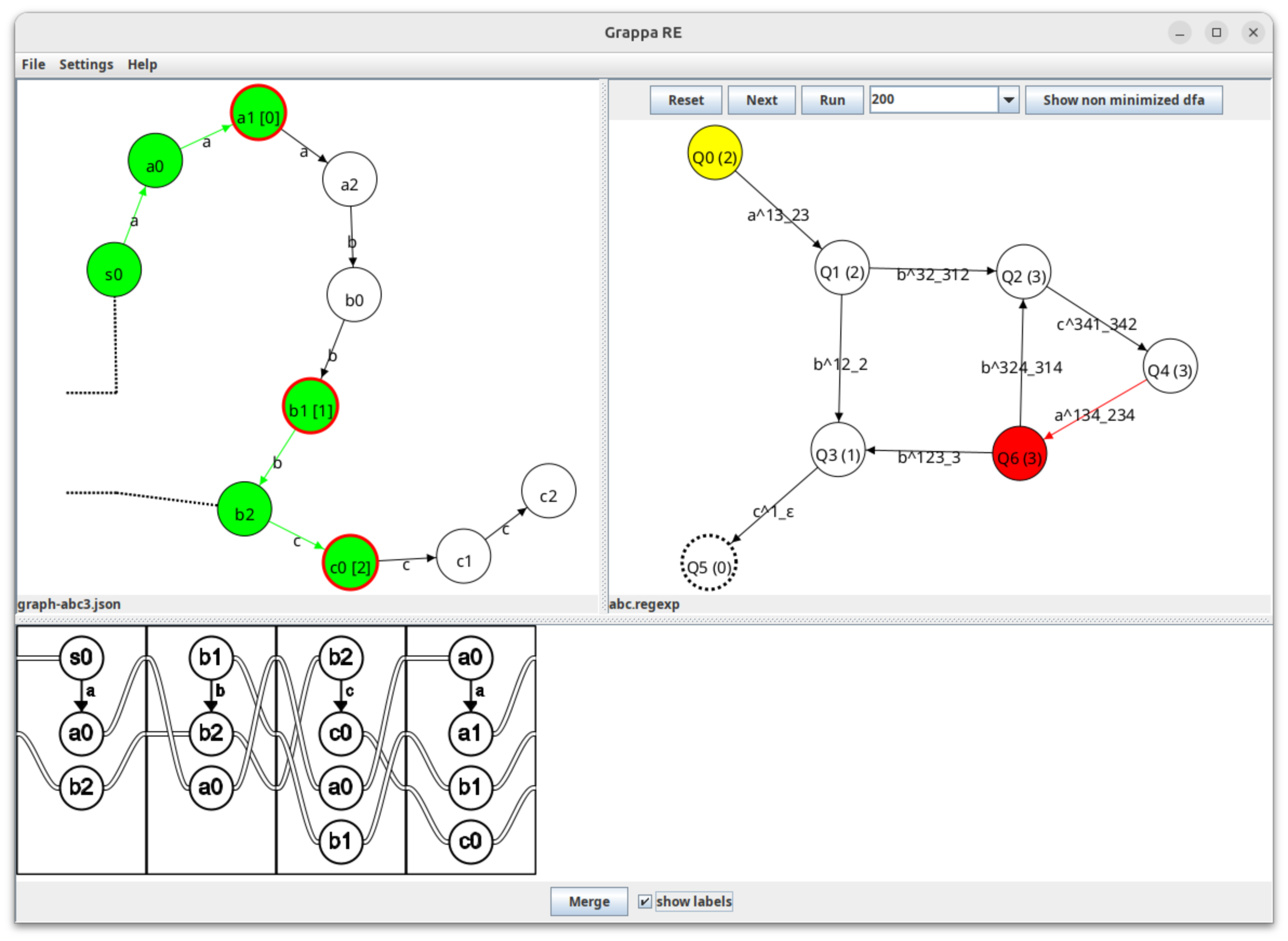}
	\\[-26mm]
	\hspace*{\fill}\includegraphics[width=.7\textwidth]{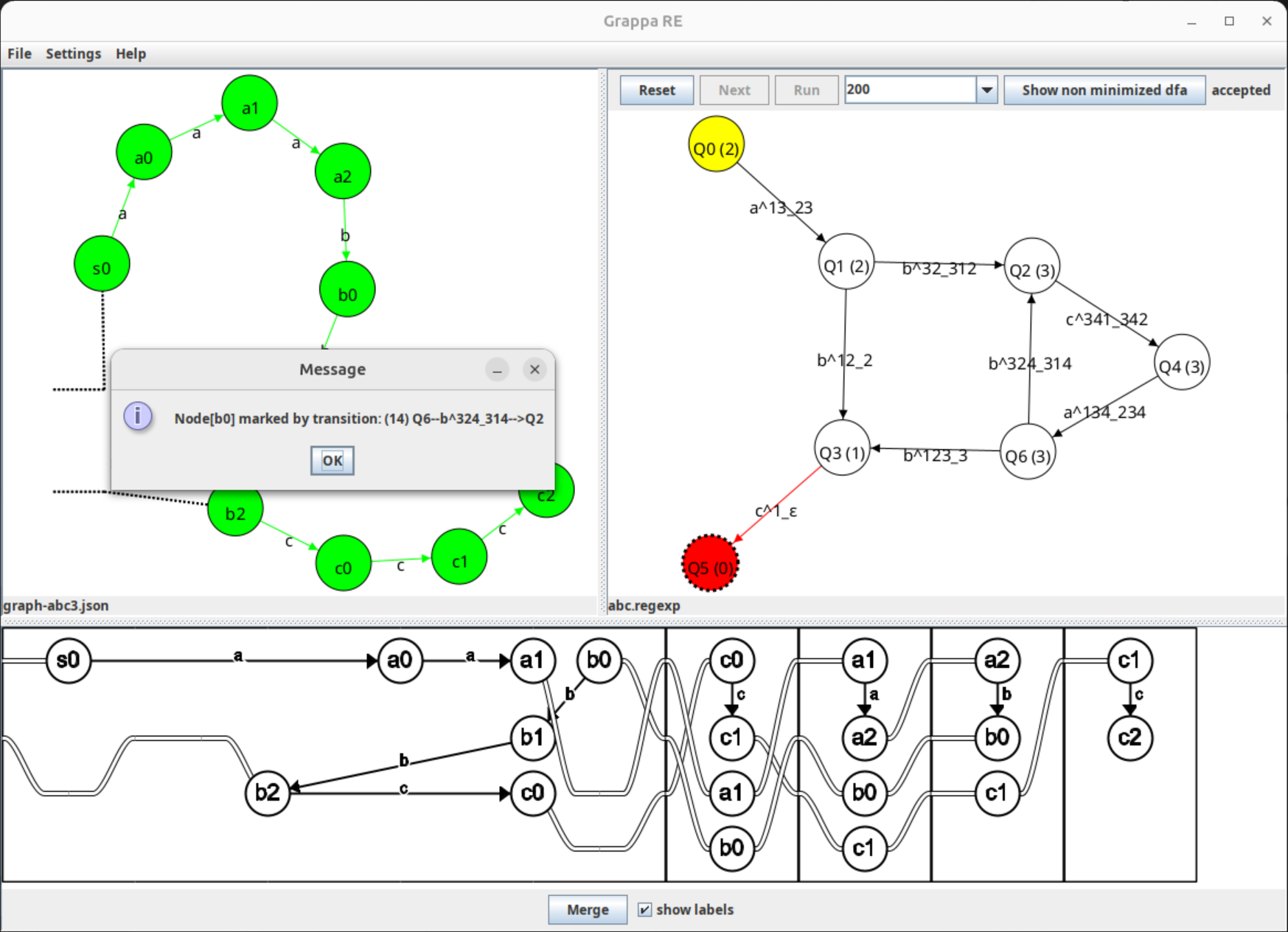}
	\caption{Two screenshots of \GrappaRE with a graph representing $aaabbbccc$
		and the automaton recognizing $a^nb^nc^n$ (see also
		\figref{f:abc-auto}). Top: The current state sequence: Q0, Q1, Q2, Q4
		and Q6. Lower: Completed execution (accepted graph). A dialog tells you
		which transition has read and marked a node by clicking on it, here
		\textit{b0}. At the bottom you can see the result after the user merged
		the first four atoms by clicking the \emph{Merge} button four times.}
	\label{f:gui}
\end{figure}
\GrappaRE also provides a GUI that allows the user to interactively visualize
the recognition process. Specifically, as shown in \figref{f:gui}, the \GrappaRE
window is divided into three parts: on the top left is a visualization of the
input graph, on the top right is the visualization of the minimized DFA and a
bar to control execution, and at the bottom is the sequence of atoms recognized
during execution. Since the input graph does not provide any layout information,
a force-directed layout algorithm~\cite{GraphStream} is used to display it,
which also allows the user to manually move nodes by dragging them with the
mouse. The same type of interface was used for the DFA, where the initial state
is highlighted in yellow and accepting states have a dashed border instead.

The user can advance the execution of the DFA manually step by step by pressing
the \textit{Next} button, or automatically at a selected time interval by
pressing the \textit{Run} button. During this operation, the current state of
the DFA is highlighted in red, as well as the edges representing the transitions
when they are followed. ``Consumed'' input graph elements are also highlighted
(in green) during execution, and it is also possible to click on them later to
get information about which transition marked them (as shown in \figref{f:gui}
below).

Finally, at the bottom, the sequence of atoms is shown as it was detected during
recognition. For ease of understanding, the user can also click the \emph{Merge}
button to replace the first two atoms by their concatenation. This merging
process is even animated. It is possible to keep clicking this button to
iteratively merge the two leftmost graphs until the bottom contains only one
graph, i.e., the concatenation of all the atoms at the beginning. The user can
also save this visualization as a graphic file in SVG format.

%
\section{Examples and Experiments}%
\label{s:experiments}
To confirm that graph recognition with finite automata can indeed run in linear
time as claimed in \sectref{s:graphs} and \cite{drewes-hoffmann-minas:23a}, the
recognition time was measured for graphs of four different graph languages. The
graph languages are $a^nb^nc^n$ and \Spikes, as introduced in
\sectsref{s:graphs} and~\ref{s:grappa-re}, respectively, and \Pali and \Wheels.
\Pali contains all graphs representing palindromes over $\{a,b\}$, and \Wheels
contains all wheel graphs as defined in \cite[p.~92]{Habel:92} and shown in
\figref{f:wheel}. The latter two are defined by %
\begin{figure}[tb]
	\centering
	\begin{tikzpicture}[x=10mm,y=10mm]
		\node[o,thick] (0) at (0,0) {};
		\node[o,thick] (1) at (1,0) {};
		\node[o,thick] (2) at (0.5,0.866) {};
		\node[o,thick] (3) at (-0.5,0.866) {};
		\node[o,thick] (4) at (-1,0) {};
		\node[o,thick] (5) at (-0.5,-0.866) {};
		\node[o,thick] (6) at (0.5,-0.866) {};
		\path
		(0) edge[->] node[above=-0.5mm] {\small$s$} (1)
		(0) edge[->] node[above left=-1.5mm] {\small$s$} (2)
		(0) edge[->] node[left=-0.5mm] {\small$s$} (3)
		(0) edge[->] node[below=-0.5mm] {\small$s$} (4)
		(0) edge[->] node[below right=-1.5mm] {\small$s$} (5)
		(0) edge[->] node[right=-0.5mm] {\small$s$} (6)
		(1) edge[->,bend right=20] node[right] {\small$t$} (2)
		(2) edge[->,bend right=20] node[above=-0.5mm] {\small$t$} (3)
		(3) edge[->,bend right=20] node[left] {\small$t$} (4)
		(4) edge[->,bend right=20] node[left] {\small$t$} (5)
		(5) edge[->,bend right=20] node[below=-0.5mm] {\small$t$} (6)
		(6) edge[->,bend right=20] node[right] {\small$t$} (1)
		;
	\end{tikzpicture}
	\vskip-2mm
	\caption{A wheel graph with six spokes.}
	\label{f:wheel}
\end{figure}
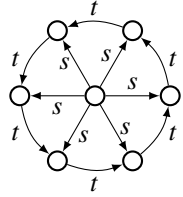%
\begin{align*}
	 & (a^{13}_{23} \; a^{32}_{31} \mid b^{13}_{23} \;
	b^{32}_{31})^\ast \;
	(a^{13}_{23} \; a^{12}_\emptyseq \mid b^{13}_{23} \;
	b^{12}_\emptyseq \mid a^{12}_\emptyseq \mid b^{12}_\emptyseq)
	 &                                                 & \text{Palindromes} \\
	 & t^\emptyseq_{12} \; s^{32}_{123} \;
	( t^{314}_{324} \; s^{123}_{123} )^\ast \;
	t^{312}_{32} \; s^{12}_\emptyseq
	 &                                                 & \text{Wheels}
\end{align*}

The experiments consisted of running graph recognition using both the simple and
the efficient edge selection algorithms (see
\sectref{sec:EfficientEdgeSelection}) on input graphs of increasing size and
measuring the execution time of the recognition algorithm. They were run on a
2022 Mac Studio with an Apple M1 Max processor, 64 GiB of RAM, and the Temurin
21 \Java virtual machine. To get more accurate measurements, each recognition
run was repeated 40 times on the same graph, shuffling the edges randomly each
time to avoid any bias due to favorable or unfavorable ordering of the edges.
The average execution time was then calculated by excluding the four worst times
to exclude cases that were slowed down by garbage collection. The results of
such executions are shown in \figsref{f:expABC} and~\ref{f:expSpike} for the
languages $a^nb^nc^n$ and \Spikes, respectively. In the plots, the x-axis shows
the number of edges of the input graph and the y-axis shows the execution time
in seconds. They clearly show that the runtime for the simple implementation of
edge selection is quadratic and that the efficient implementation is indeed much
faster.%
\begin{figure}[b]
	\newcommand\myscale{0.75}
	\begin{minipage}[t]{0.47\textwidth}
		\begin{tikzpicture}[transform shape,scale=\myscale]
			\begin{axis}[
					ylabel=Running time (seconds),
					xlabel=Graph size (number of edges),
					legend pos=north west,
					width=0.95\textwidth/\myscale,
					height=0.7\textwidth/\myscale,
					grid=major,
					xmin=0,
					ymin=0,
					xmax=25023,
					ymax=1,
					xtick distance=5000,
					ytick distance=0.2,
					xticklabel style={/pgf/number format/precision=1,
							/pgf/number format/fixed,
							/pgf/number format/fixed zerofill},
					yticklabel style={/pgf/number format/precision=1,
							/pgf/number format/fixed,
							/pgf/number format/fixed zerofill},
				]
				\addplot+[smooth] coordinates {
						(3,0.000044)
						(1254,0.002573)
						(2505,0.009262)
						(3756,0.02262)
						(5007,0.040294)
						(6258,0.061861)
						(7509,0.088394)
						(8760,0.118662)
						(10011,0.151296)
						(11262,0.191438)
						(12513,0.236669)
						(13764,0.287177)
						(15015,0.344987)
						(16266,0.402106)
						(17517,0.470022)
						(18768,0.538338)
						(20019,0.613789)
						(21270,0.698013)
						(22521,0.784312)
						(23772,0.879988)
						(25023,0.980046)
					};
				\addlegendentry{simple}
				\addplot+[smooth] coordinates {
						(3,0.000254)
						(1254,0.00132)
						(2505,0.001469)
						(3756,0.001596)
						(5007,0.001968)
						(6258,0.002398)
						(7509,0.002894)
						(8760,0.003413)
						(10011,0.00395)
						(11262,0.004606)
						(12513,0.005384)
						(13764,0.006144)
						(15015,0.006926)
						(16266,0.007769)
						(17517,0.008648)
						(18768,0.009519)
						(20019,0.010439)
						(21270,0.011424)
						(22521,0.012395)
						(23772,0.013307)
						(25023,0.013999)
					};
				\addlegendentry{efficient}
			\end{axis}
		\end{tikzpicture}
		\caption{Recognizing $a^nb^nc^n$ graphs using the simple and the efficient implementation.}\label{f:expABC}
	\end{minipage}
	\hfill
	\begin{minipage}[t]{0.47\textwidth}
		\begin{tikzpicture}[transform shape,scale=\myscale]
			\begin{axis}[
					ylabel=Running time (seconds),
					xlabel=Graph size (number of edges),
					legend pos=north west,
					width=0.95\textwidth/\myscale,
					height=0.7\textwidth/\myscale,
					grid=major,
					xmin=0,
					ymin=0,
					xmax=25001,
					ymax=1,
					xtick distance=5000,
					ytick distance=0.2,
					xticklabel style={/pgf/number format/precision=1,
							/pgf/number format/fixed,
							/pgf/number format/fixed zerofill},
					yticklabel style={/pgf/number format/precision=1,
							/pgf/number format/fixed,
							/pgf/number format/fixed zerofill},
				]

				\addplot+[smooth] coordinates {
						(1,0.000032)
						(1251,0.014885)
						(2501,0.068094)
						(3751,0.159003)
						(5001,0.285528)
						(6251,0.436932)
						(7501,0.61669)
						(8751,0.814664)
						(10001,1.027908)
						(11251,1.36845)
						(12501,1.738273)
						(13751,2.129065)
						(15001,2.530512)
					};
				\addlegendentry{simple}
				\addplot+[smooth] coordinates {
						(1,0.000115)
						(1251,0.001326)
						(2501,0.00189)
						(3751,0.002696)
						(5001,0.00366)
						(6251,0.00476)
						(7501,0.005895)
						(8751,0.007189)
						(10001,0.0086)
						(11251,0.009957)
						(12501,0.011718)
						(13751,0.01322)
						(15001,0.014675)
						(16251,0.016122)
						(17501,0.017692)
						(18751,0.01923)
						(20001,0.020849)
						(21251,0.022467)
						(22501,0.024063)
						(23751,0.025782)
						(25001,0.027941)
					};
				\addlegendentry{efficient}
			\end{axis}
		\end{tikzpicture}
		\caption{Recognizing \Spikes graphs using the simple and the efficient implementation.}\label{f:expSpike}
	\end{minipage}
\end{figure}

\begin{figure}[tb]
	\newcommand\myscale{0.75}
	\centering
	\begin{tikzpicture}[transform shape,scale=\myscale]
		\begin{axis}[
				ylabel=Running time (seconds),
				xlabel=Graph size (number of edges),
				legend pos=north west,
				width=9cm/\myscale,
				height=6cm/\myscale,
				grid=major,
				xmin=0,
				ymin=0,
				xmax=1000023,
				ymax=2.5,
				xtick distance=200000,
				ytick distance=0.5,
				xticklabel style={/pgf/number format/precision=1,
						/pgf/number format/fixed,
						/pgf/number format/fixed zerofill},
				yticklabel style={/pgf/number format/precision=1,
						/pgf/number format/fixed,
						/pgf/number format/fixed zerofill},
			]
			\addplot[only marks,mark=pentagon*,mark size=2.7pt,blue!70!black] coordinates {
					(1,0.000249)
					(50001,0.068293)
					(100001,0.162474)
					(150001,0.259836)
					(200001,0.375295)
					(250001,0.477949)
					(300001,0.588746)
					(350001,0.70144)
					(400001,0.832813)
					(450001,0.945866)
					(500001,1.045365)
					(550001,1.166592)
					(600001,1.291738)
					(650001,1.407121)
					(700001,1.526145)
					(750001,1.637952)
					(800001,1.740331)
					(850001,1.87908)
					(900001,2.016661)
					(950001,2.14553)
					(1000001,2.249376)
				};
			\addlegendentry{Spikes}
			\addplot[only marks,mark=square*,mark size=2.6pt,red!80!black] coordinates {
					(2,0.000141)
					(50002,0.048596)
					(100002,0.122511)
					(150002,0.198774)
					(200002,0.284676)
					(250002,0.364166)
					(300002,0.446361)
					(350002,0.528617)
					(400002,0.607965)
					(450002,0.690733)
					(500002,0.775979)
					(550002,0.863497)
					(600002,0.944006)
					(650002,1.031202)
					(700002,1.118537)
					(750002,1.207114)
					(800002,1.30056)
					(850002,1.388135)
					(900002,1.478079)
					(950002,1.572478)
					(1000002,1.653589)
				};
			\addlegendentry{Palindromes}
			\addplot[only marks,mark=*,mark size=2.6pt,green!40!black] coordinates {
					(6,0.000098)
					(50006,0.03784)
					(100006,0.097194)
					(150006,0.162721)
					(200006,0.234334)
					(250006,0.306356)
					(300006,0.378794)
					(350006,0.453502)
					(400006,0.532581)
					(450006,0.606402)
					(500006,0.688198)
					(550006,0.767363)
					(600006,0.845857)
					(650006,0.928278)
					(700006,0.995972)
					(750006,1.081507)
					(800006,1.15358)
					(850006,1.24536)
					(900006,1.319723)
					(950006,1.403734)
					(1000006,1.494549)
				};
			\addlegendentry{Wheels}
			\addplot[only marks,mark=triangle*,mark size=3.3pt,red!60!black] coordinates {
					(3,0.000288)
					(50004,0.038148)
					(100005,0.097446)
					(150006,0.160519)
					(200007,0.22866)
					(250008,0.295887)
					(300009,0.36562)
					(350010,0.43637)
					(400011,0.511611)
					(450012,0.587116)
					(500013,0.660957)
					(550014,0.733778)
					(600015,0.804288)
					(650016,0.870255)
					(700017,0.949831)
					(750018,1.028292)
					(800019,1.105595)
					(850020,1.179626)
					(900021,1.268966)
					(950022,1.330447)
					(1000023,1.389782)
				};
			\addlegendentry{$a^nb^nc^n$}
			\addplot[no markers,solid,semithick,blue!70!black] coordinates {
					(1,0.000249)
					(50001,0.068293)
					(100001,0.162474)
					(150001,0.259836)
					(250001,0.477949)
					(1000001,2.249376)
				};
			\addplot[no markers,solid,semithick,red!80!black] coordinates {
					(2,0.000141)
					(50002,0.048596)
					(100002,0.122511)
					(150002,0.198774)
					(200002,0.284676)
					(250002,0.364166)
					(1000002,1.653589)
				};
			\addplot[no markers,solid,semithick,green!40!black] coordinates {
					(6,0.000098)
					(50006,0.03784)
					(100006,0.097194)
					(150006,0.162721)
					(200006,0.234334)
					(250006,0.306356)
					(300006,0.378794)
					(1000006,1.494549)
				};
			\addplot[no markers,solid,semithick,red!60!black] coordinates {
					(3,0.000288)
					(50004,0.038148)
					(100005,0.097446)
					(250008,0.295887)
					(1000023,1.389782)
				};
		\end{axis}
	\end{tikzpicture}
	\caption{Recognizing different kinds of graphs using the efficient implementation.}\label{f:comparison}
\end{figure}
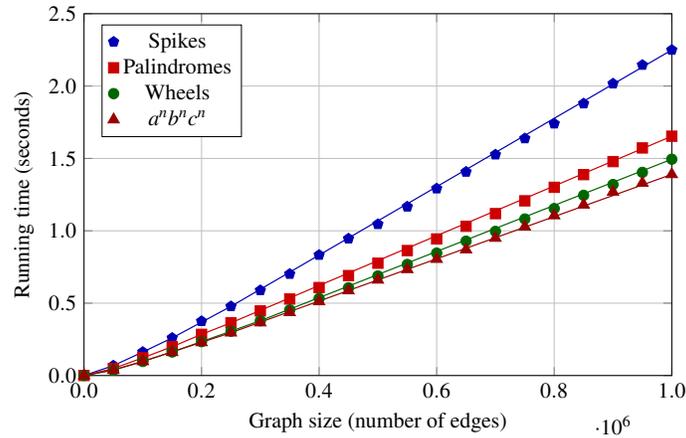

In addition, \figref{f:comparison} compares the time taken to recognize graphs
using only the efficient edge selection algorithm for the four languages on
larger input graphs; in this case, we reduced the number of runs of each test
from 40 to 6 in order to reduce the time required given the larger size of the
graphs. It can be seen that the increase in execution time remains linear even
when scaling up to large graphs with a million edges.%

In addition to the recognition execution time, we also measured the time taken
for the various operations required to create the DFA for the four languages.
The operations of generating the automaton from a regular expression, checking
the automaton for ambiguity, performing the powerset construction, minimizing
the DFA, and checking the DFA for the FEC property took less than \ms{1} in
total. Checking the TS property (and reordering the transitions) took on average
\ms{7.5} (between \ms{6.8} and \ms{8.3}). Since it is also possible to store the
DFA resulting from these operations and reuse it for recognition operations, the
time taken by these operations can be considered negligible.


Regarding the times shown in the graphs, in the case of the efficient edge
selection implementation, they include both the time to generate the optimized
data structures of the input graph and the time to execute the DFA. In
particular, the time to create the optimized data structures was found to be
about two-thirds of the total time, while the time to execute the DFA was about
one-third of the total time.

%
\section{Conclusions}%
\label{s:concl}

This paper has described \GrappaRE, a tool for analyzing graphs in the sense
that it decides for a given graph whether it can be generated by a given finite
automaton or not. \GrappaRE implements the approach recently proposed by
Hoffmann, Drewes, and Minas~\cite{drewes-hoffmann-minas:23a}, where graphs are
considered as interpretations of strings and finite automata are used to define
a graph language and also as a device for deciding the membership problem. In
particular, \GrappaRE implements the procedures for checking whether a given
automaton allows efficient graph recognition. This paper has shown that the
recognition is indeed linear in the size of the input graph, and it has
described the implementation approach that was necessary to achieve linear
runtime complexity. Experiments have shown that even very large graphs can be
analyzed very quickly with this implementation, since only about
$2\,\mu\textrm{s}$ is needed to analyze each edge of the input graph on standard
computers.

Future work will consider applications of the approach 
described in
\cite{drewes-hoffmann-minas:23a}, e.g., by using regular expressions to be used
in the nested graph conditions of Habel and Pennemann
\cite{Habel-Pennemann:05,Pennemann:09} to specify global graph properties
and check them with automata.

\bibliographystyle{eptcs}
\bibliography{References}
\end{document}